\documentclass[preprint,10pt]{elsarticle}

\usepackage{hyperref}
\usepackage{subfig}
\usepackage{mathtools}
\usepackage{comment}
\usepackage{booktabs}
\usepackage{amssymb}
\usepackage{url}
\usepackage{float}
\usepackage{tabularx}
\usepackage{rotating}
\usepackage{enumerate}
\usepackage{array}
\usepackage{xspace}

\usepackage{amsthm}
\usepackage{amsmath}
\usepackage{algorithm}
\usepackage[noend]{algpseudocode}
\usepackage{booktabs}

\usepackage[a4paper,left=3cm,right=3cm,top=2.5cm,bottom=2.5cm]{geometry}

\DeclareMathOperator*{\argmin}{arg\,min}

\newcommand{\fig}{\text{Fig.~}}
\newcommand{\tab}{\text{Tab.~}}
\newcommand{\eq}{\text{Eq.~}}
\newcommand{\sez}{\text{Sec.~}}

\DeclareFontEncoding{LS1}{}{}
\DeclareFontSubstitution{LS1}{stix}{m}{n}
\DeclareSymbolFont{symbolsstix}{LS1}{stixscr}{m}{n}
\SetSymbolFont{symbolsstix}{bold}{LS1}{stixscr}{b}{n}
\DeclareMathSymbol{\xx}{0}{symbolsstix}{'170}
\DeclareMathSymbol{\hh}{0}{symbolsstix}{'150}
\DeclareMathSymbol{\qq}{0}{symbolsstix}{'161}

\begin{document}

\begin{frontmatter}
\title{Neural Markov chain Monte Carlo: \\Bayesian inversion
via normalizing flows and variational autoencoders}

\author[1]{Giacomo~Bottacini}

\author[2]{Matteo~Torzoni\corref{cor1}}

\author[1]{Andrea~Manzoni}

\affiliation[1]{organization={MOX -- Department of
Mathematics, Politecnico di Milano},
city={Milan},
postcode={20133},
country={Italy}}

\affiliation[2]{organization={Department of Civil and Environmental Engineering, Politecnico di Milano}, 
city={Milan},
postcode={20133},
country={Italy}}

\cortext[cor1]{Corresponding author: \texttt{matteo.torzoni@polimi.it}}


\begin{abstract}
This paper introduces a Bayesian framework that combines Markov chain Monte Carlo (MCMC) sampling, dimensionality reduction, and neural density estimation to efficiently handle inverse problems that {\em (i)} must be solved multiple times, and {\em (ii)} are characterized by intractable or unavailable likelihood functions. The posterior probability distribution over quantities of interest is estimated via differential evolution Metropolis sampling, empowered by learnable mappings. First, a variational autoencoder performs probabilistic feature extraction from observational data. The resulting latent structure inherently quantifies uncertainty, capturing deviations between the actual data-generating process and the training data distribution. At each step of the MCMC random walk, the algorithm jointly samples from the data-informed latent distribution and the space of parameters to be inferred. These samples are fed into a neural likelihood estimator based on normalizing flows, specifically real-valued non-volume preserving transformations. The scaling and translation functions of the affine coupling layers are modeled by neural networks conditioned on the unknown parameters, allowing the representation of arbitrary observation likelihoods. The proposed methodology is validated on two case studies: {\em (i)} structural health monitoring of a railway bridge for damage detection, localization, and quantification, and {\em (ii)} estimation of the conductivity field in a steady-state Darcy’s groundwater flow problem. The results demonstrate the efficiency of the inference strategy, while ensuring that model-reality mismatches do not yield overconfident, yet inaccurate, estimates. 
\end{abstract}


\end{frontmatter}

\section{Introduction}

When analyzing physical systems, key parameters of interest are often difficult or impossible to measure directly. Instead, a forward model, derived from physical laws or empirical relationships, is typically available to describe how hidden parameters and internal states generate observables. This setup allows the estimation of unknown parameters from measured data to be cast as an inverse problem, which is often ill-posed and computationally challenging due to non-uniqueness and sensitivity to noise. 

A principled approach to address this class of problems is provided by Bayesian inference~\cite{stuart2010inverse, tarantola2004inverse}, which treats unknown parameters as random variables and frames the inverse problem in a probabilistic setting. This formulation inherently enables uncertainty quantification (UQ) and provides a natural mechanism to regularize ill-posedness. In essence, Bayesian inference updates prior beliefs by conditioning them on observational data, yielding a posterior distribution over the parameters of interest~\cite{doucet2009tutorial, doucet2001sequential}. A critical component of this process is the likelihood function, which encapsulates the assumed statistical relationship between model predictions and observations.

However, scientific and engineering applications are increasingly characterized by high-di\-men\-sio\-nal and highly complex observational data, as seen in fields such as genomics~\cite{way2018extracting}, imaging~\cite{pu2016vae_images_labels_captions,ehrhardt2022autoencoders}, video analysis~\cite{fan2020video}, and natural language processing~\cite{kusner2017grammar}. High-dimensional data pose several challenges, including overfitting and reduced generalization, which can compromise the performance and robustness of downstream inference models. Additionally, when dealing with high-dimensional observables~\cite{arridge2019solving} or a large number of parameters to be inferred (e.g., spatially distributed material properties)~\cite{iglesias2013evaluation}, the parameter-to-observation map may exhibit low sensitivity, meaning that significant variations in parameters produce only subtle changes in the outputs. This results in ill-posedness, poorly identifiable solutions, and elevated epistemic uncertainty. Moreover, accurately characterizing the posterior distribution often requires a large number of forward evaluations, which can be computationally prohibitive when each evaluation involves expensive simulations  (e.g., finite element analyses). Finally, in many real-world applications, the likelihood function may be intractable, unavailable in closed form, or too costly to evaluate directly.
 
These challenges have motivated the integration of recent advances in machine learning, particularly deep learning (DL), into Bayesian inference pipelines to enhance computational scalability~\cite{antil2021deepnn,deveney2019deepsurrogate,yan2019adaptive}. A prominent direction in this context involves dimensionality reduction~\cite{goh2019uqvae}, through the projection of data into a lower-dimensional space. This approach not only reduces computational costs but can also help uncover the intrinsic structure of the underlying generative process. A wealth of dimensionality reduction techniques have been proposed in the literature~\cite{akkarapatty2016dimensionality}, and they can generally be categorized into two classes: {\em (i)} feature selection and ({\em ii)} feature extraction. The former retains a subset of the original variables, whereas the latter maps the data into a new space through learned or predefined transformations. Recent research has increasingly focused on learnable feature extraction~\cite{mutlag2020feature}, aiming to automatically uncover hierarchical latent patterns without relying on manual variable selection or domain-specific assumptions.

Traditional feature extraction methods rely on projection techniques that map high-dimensional data onto lower-dimensional subspaces while seeking to minimize information loss. Among the most widely used approaches are principal component analysis (PCA) and linear discriminant analysis (LDA)~\cite{xanthopoulos2013lda}. PCA identifies orthogonal directions, or principal components, that capture the maximum variance in the data, providing an unsupervised linear embedding. In contrast, LDA is a supervised method that seeks to optimize class separability by maximizing inter-class variance while minimizing intra-class variance within the projection subspace. Despite their interpretability and computational efficiency, both PCA and LDA are inherently restricted to linear transformations, which limits their ability to model complex, nonlinear relationships. 

In this context, autoencoders (AEs) provide a nonlinear DL-based alternative for compressing high-dimensional observations into compact latent representations that capture the most informative features of the data. Extensions of the standard AE structure include sparse AEs, denoising AEs, and variational autoencoders (VAEs)~\cite{doersch2016tutorial,book:DL_book}. Variational autoencoders, in particular, have demonstrated superior capabilities in quantifying uncertainty in learned representations -- a crucial requirement in applications involving decision-making under uncertainty. Specifically, they impose a probabilistic encoding scheme in which the latent representation not only supports input reconstruction but is also constrained to follow a predefined prior distribution. This probabilistic regularization promotes a smooth, continuous, and semantically meaningful latent space, thereby enabling the generation of plausible samples via latent space interpolation or stochastic sampling.

From a complementary perspective, the emergence of simulation-based inference~\cite{inference,papamakarios2019neural,papamakarios2019snl} offers a framework for performing Bayesian inference when the likelihood function is intractable or unavailable. These methods bypass explicit likelihood evaluation by leveraging generative or surrogate models, such as neural density estimators~\cite{papamakarios2021normalizing}, to learn the statistical relationship between parameters and data directly from observations.

Several surrogate-based strategies have been proposed to accelerate Bayesian inverse uncertainty quantification. Early works, see~\cite{marzouk2009}, have employed polynomial chaos expansions to approximate forward models or likelihood functions, enabling efficient Markov chain Monte Carlo (MCMC) sampling in PDE-constrained inverse problems. Other approaches reduce or bypass sampling by constructing global surrogates of the inverse map or the posterior distribution~\cite{sudret2016}. More recently, surrogate models have been integrated with advanced MCMC schemes to achieve substantial computational savings while preserving Bayesian consistency~\cite{rossat2022,reviewGJI}. Despite these advances, handling high-dimensional observations while preserving sampling-based uncertainty quantification remains a major challenge.

In this work, we propose \textit{neural} MCMC sampling, a likelihood-free approach~\cite{sisson2011likelihoodfree, hermans2020likelihoodfree} that integrates two DL components to enable efficient inference of parameters of interest. The first component is a VAE, trained to extract informative features from high-dimensional observations while explicitly quantifying uncertainty. At each MCMC iteration, a new parameter proposal is generated and paired with a latent code sampled from the VAE encoding. This pair is then passed to a second generative model for density estimation. This second component is a conditional normalizing flow architecture~\cite{winkler2019conditional}, trained to approximate the likelihood function using real-valued non-volume preserving (RealNVP) transformations~\cite{dinh2017realnvp}. The output of this surrogate likelihood model enables direct computation of the MCMC acceptance ratio, thereby efficiently guiding the sampling process. This hybrid framework enables scalable posterior inference in high-dimensional and computationally intensive inverse problems, while also providing uncertainty quantification through the learned variational structure.

The paper is organized as follows. Section~\ref{sec:2} describes the proposed computational methodology. Section~\ref{sec:3} presents numerical results for two case studies: (i) a structural health monitoring scenario involving the detection, localization, and quantification of damage in a railway bridge; and
(ii) a steady-state groundwater flow problem governed by Darcy’s law, with a focus on estimating a spatially varying hydraulic conductivity field. Finally, Section~\ref{sec:conclusion} draws some conclusions.

\section{Data assimilation via normalizing flows and variational autoencoders}
\label{sec:2}
This section details the proposed simulation-based inference framework. The approach is built upon two components tightly integrated within an MCMC algorithm: (i) a modified VAE for probabilistic feature extraction from observational data, and (ii) a neural likelihood estimator based on a conditional normalizing flow with coupling layers parameterized by neural networks conditioned on the parameters to be inferred. The VAE enables low-dimensional and noise-robust representation of high-dimensional observations, while the conditional normalizing flow provides a flexible likelihood approximation in the latent space, allowing for accurate UQ within a Bayesian setting.

The resulting MCMC scheme jointly samples from the data-informed latent distribution and the parameter space to infer the posterior of target parameters. In the following, section~\ref{sec:MCMC} outlines the overall framework and explains how the two neural components are embedded within the inference procedure. Sections~\ref{sec:NF} and~\ref{sec:VAE} then describe the architectures and training procedures of the RealNVP and the VAE, respectively.

\subsection{Simulation-based inference scheme}
\label{sec:MCMC}
We formulate the inverse problem in a probabilistic setting. Adopting a Bayesian perspective, we treat $N_\text{par}$ target unknown parameters $\boldsymbol{\lambda}\in\mathbb{R}^{N_\text{par}}$ as random variables to be inferred from data. The prior distribution $p(\boldsymbol{\lambda})$, representing the initial belief about the parameters before assimilating any data, is updated through the likelihood function $p(\mathbf{h}\mid\boldsymbol{\lambda})$ upon observing data $\mathbf{h}\in\mathbb{R}^{N_h}$. By Bayes' theorem, the posterior distribution is given by:
\begin{equation}
    p(\boldsymbol{\lambda} \mid \mathbf{h}) = \frac{p(\mathbf{h} \mid \boldsymbol{\lambda}) p(\boldsymbol{\lambda})}{\int p(\mathbf{h} \mid \boldsymbol{\lambda}) p(\boldsymbol{\lambda}) \, d\boldsymbol{\lambda}}.
\end{equation}

In many applications, evaluating the posterior $p(\boldsymbol{\lambda} \mid \mathbf{h})$ in closed form is infeasible due to both the denominator $p(\mathbf{h})=\int p(\mathbf{h} \mid \boldsymbol{\lambda}) p(\boldsymbol{\lambda}) \, d\boldsymbol{\lambda}$, which involves an integral over the entire parameter space that may be high-dimensional, and the need for a complex forward model to compute $p(\mathbf{h} \mid \boldsymbol{\lambda})$, making the likelihood function available only numerically. As a result, numerical sampling methods are typically employed. 

Monte Carlo methods provide a practical approach for approximating the posterior distribution by drawing samples from $p(\boldsymbol{\lambda} \mid \mathbf{h})$. Several algorithms exist for sampling the posterior, with MCMC methods being among the most widely used~\cite{art:AM_Green,gilks1996markov,robert1999monte}. A notable example is the Metropolis-Hastings algorithm~\cite{art:MH}, which constructs a Markov chain by iteratively proposing candidate states $\boldsymbol{\lambda}'$ from a proposal distribution. To obtain the $n$-th state of the chain, with $n=1,\ldots,N_\text{s}$ and $N_\text{s}$ denoting the chain length, proposals are drawn from $q_\text{MH}(\boldsymbol{\lambda}'\mid\boldsymbol{\lambda}^{(n-1)})$. This proposal is typically chosen as a Gaussian density centered at the current state $\boldsymbol{\lambda}^{(n-1)}$ with covariance matrix $\boldsymbol{\Sigma}\in\mathbb{R}^{N_\text{par}\times N_\text{par}}$, i.e. $\boldsymbol{\lambda}' \sim \mathcal{N}(\boldsymbol{\lambda}^{(n-1)}, \boldsymbol{\Sigma})$, thereby generating local, symmetric perturbations around the current state.

While Gaussian proposals are widely applicable, they often struggle in high-dimensional or strongly correlated parameter spaces. For this reason, we adopt a more robust and adaptive strategy based on differential evolution MCMC (DE-MCMC)~\cite{terbraak2006demc}. This family of sampling algorithms employs proposal mechanisms that combine information from past states of the chain to capture parameter correlations and promote global exploration of the parameter space. The standard DE-MCMC proposal takes the following form:
\begin{equation}
\boldsymbol{\lambda}' = \boldsymbol{\lambda}^{(n-1)} + \gamma (\mathbf{\Lambda}_{R1} - \mathbf{\Lambda}_{R2}) + \boldsymbol{\epsilon},
\label{eq:DE-MCMC}
\end{equation}
where $\mathbf{\Lambda}_{R1},\mathbf{\Lambda}_{R2}\in\mathbb{R}^{N_\text{par}}$ are two distinct samples drawn from an archive (or population) of past states $\mathbf{\Lambda}$. This archive collects previously accepted states at the time of generating $\boldsymbol{\lambda}^{(n)}$ and is systematically updated with new $\boldsymbol{\lambda}$ states at intervals of $k_T\in\mathbb{N}$ iterations. The parameter $\gamma\in\mathbb{R}^+$ is a non-negative scaling factor that controls the jump size, and $\boldsymbol{\epsilon}\in\mathbb{R}^{N_\text{par}}$ is a small random perturbation introduced to ensure ergodicity. In practice, $\gamma$ is chosen to scale inversely with the square root of the parameter dimension, with $\gamma=\frac{2.38}{\sqrt{2N_\text{par}}}$ being a recommended value that yields near-optimal acceptance rates for multivariate distributions~\cite{terbraak2006demc}.

In this work, we employ the DEMetropolisZ sampler from the PyMC library~\cite{PyMC}, an adaptive DE-MCMC algorithm that leverages an evolving archive of past states while preserving the standard Metropolis–Hastings acceptance mechanism. The probability of accepting $\boldsymbol{\lambda}'$ as the next state $\boldsymbol{\lambda}^{(n)}$ of the Markov chain targeting $p(\boldsymbol{\lambda}\mid\mathbf{h}) \propto p(\mathbf{h}\mid\boldsymbol{\lambda})p(\boldsymbol{\lambda})$ is therefore defined as:
\begin{equation}
    \alpha(\boldsymbol{\lambda}^{(n-1)}, \boldsymbol{\lambda}') = \min \left\{ 1, \frac{p(\mathbf{h}\mid\boldsymbol{\lambda}')p(\boldsymbol{\lambda}')} {p(\mathbf{h}\mid\boldsymbol{\lambda}^{(n-1)})p(\boldsymbol{\lambda}^{(n-1)}) } \right\},
\label{eq:acceptance}
\end{equation}
which corresponds to the Metropolis–Hastings acceptance probability in the case of a symmetric proposal distribution. Since the stochastic proposal rule~\eqref{eq:DE-MCMC} is symmetric by construction, this formulation guarantees that the resulting Markov chain satisfies the detailed balance condition, ensuring that its stationary distribution coincides with the target posterior~\cite{gilks1996markov}.

Evaluating $\alpha(\boldsymbol{\lambda}^{(n-1)}, \boldsymbol{\lambda}')$ only requires computing the likelihood $p(\mathbf{h} \mid \boldsymbol{\lambda}')$ and the prior $p(\boldsymbol{\lambda}')$. However, for many modern problems, directly evaluating the likelihood  $p(\mathbf{h}\mid\boldsymbol{\lambda}')$ is computationally prohibitive or even infeasible. Likelihood-free Bayesian methods~\cite{wegmann2009efficient} provide a viable strategy for addressing this limitation. Approximate Bayesian computation (ABC) is a prominent example of a likelihood-free method that bypasses explicit likelihood evaluations by accepting candidate parameter values when the simulated data are sufficiently \textit{close} to the observed data~\cite{sunnaaker2013approximate}. While conceptually straightforward, ABC can be highly inefficient, often requiring an extremely large number of simulations to obtain a reasonable approximation of the posterior distribution.

To overcome the limitations of common likelihood-free methods such as ABC, we propose a simulation-based inference framework~\cite{inference} that integrates DE-MCMC sampling with neural density estimation~\cite{papamakarios2019neural}. Specifically, we employ a conditional normalizing flow (CNF)~\cite{winkler2019conditional} as a surrogate likelihood model to approximate $p(\mathbf{h} \mid \boldsymbol{\lambda})$. Once trained, the CNF provides tractable and differentiable likelihood evaluations that can be directly incorporated into the acceptance ratio~\eqref{eq:acceptance}. This scheme is further enhanced by a VAE that first compresses the raw observations into a low-dimensional, informative latent representation. This dimensionality reduction step enables the CNF to operate on a substantially lower-dimensional input, improving the overall stability and computational efficiency. By jointly modeling (and sampling from) both the latent VAE representation and the parameter space, the proposed inference scheme allows the CNF to instantaneously estimate observation likelihoods while preserving the reliability of the inference process. 

\begin{figure}[t]
    \centering
    \includegraphics[width=\linewidth]{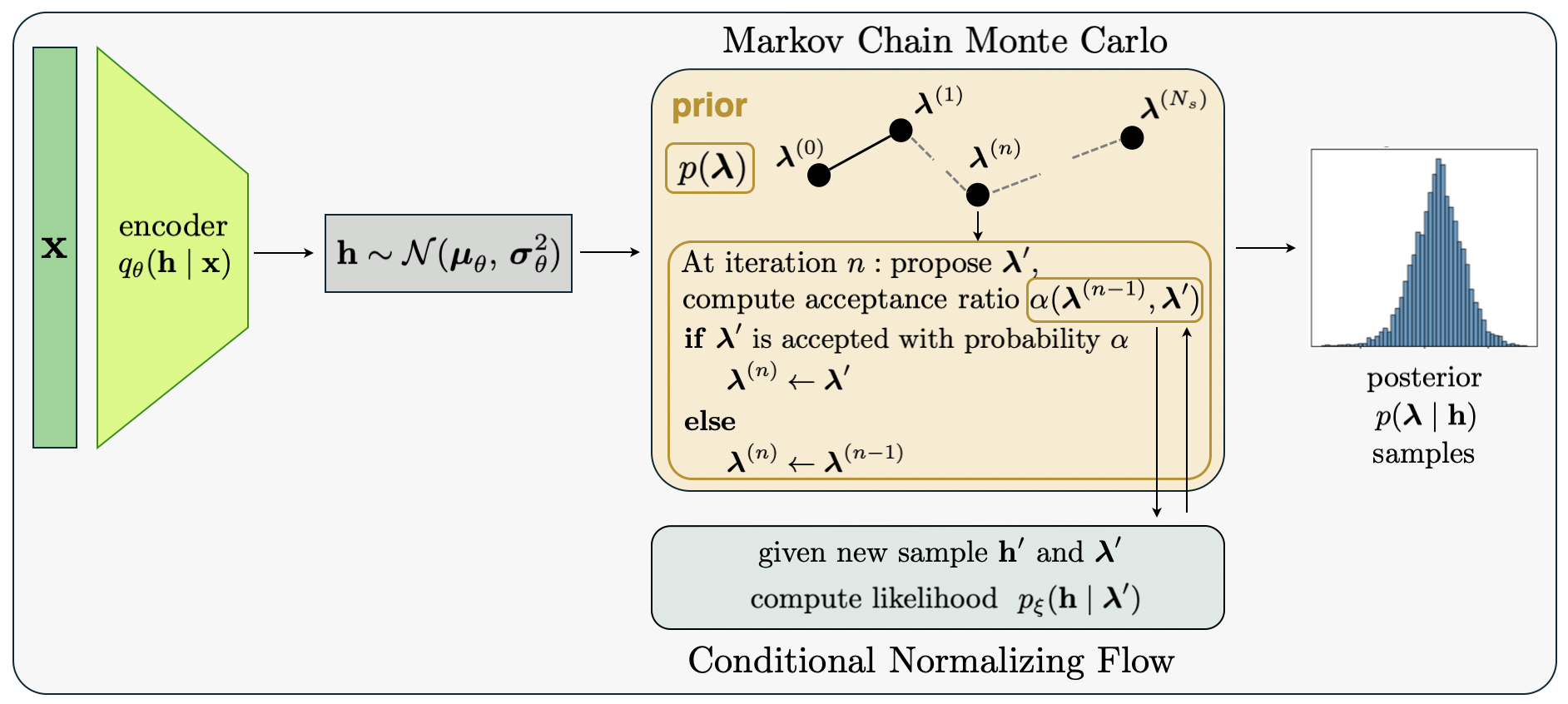}
    \caption{
    Schematic of the \textit{neural} Markov chain Monte Carlo (MCMC) sampling scheme for simulation-based inference of the parameter posterior $\mathbf{p}(\boldsymbol{\lambda}\mid\mathbf{h})$. The encoder branch of a variational autoencoder (VAE) performs probabilistic feature extraction from the observed data $\mathbf{x}$, yielding the latent distribution $\mathbf{h}\sim\mathcal{N}(\boldsymbol{\mu}_\theta,\, \boldsymbol{\sigma}^2_\theta)$. Joint samples $\langle\mathbf{h}' ,\boldsymbol{\lambda}'\rangle$ are drawn from the latent distribution and the parameter space, respectively. A conditional normalizing flow network then estimates the conditional likelihood density $p(\mathbf{h}' \mid \boldsymbol{\lambda}')$, which is used to efficiently compute the MCMC acceptance ratio $\alpha(\boldsymbol{\lambda}^{(n-1)}, \boldsymbol{\lambda}')$.}
    \label{fig:MCMC}
\end{figure}

We refer to our architecture as \textit{neural} MCMC, composed of three main components (see \fig\ref{fig:MCMC}):
($i$) an adaptive MCMC scheme that updates its proposal mechanism based on past samples, thereby improving mixing and sampling efficiency; ($ii$) a CNF-based estimator, trained to provide fast likelihood approximations during sampling, as detailed in \sez\ref{sec:NF}; and
($iii$) a VAE for probabilistic feature extraction from the observed data, as described in \sez\ref{sec:VAE}.

Given raw measurements $\mathbf{x}\in\mathbb{R}^{N_x}$ from an $N_x$-dimensional input space, with $N_h\ll N_x$, and a prior distribution $p(\boldsymbol{\lambda})$, the algorithm approximates the posterior $p(\boldsymbol{\lambda} \mid \mathbf{x})$. Raw data $\mathbf{x}$ are first encoded into an $N_h$-dimensional latent distribution using the VAE encoder \mbox{$q_\theta:\mathbb{R}^{N_x}\mapsto\mathbb{R}^{N_h}$}, as:
\begin{equation}
q_\theta(\mathbf{h} \mid \mathbf{x}) = \mathcal{N}(\boldsymbol{\mu}_\theta(\mathbf{x}),\, \boldsymbol{\sigma}^2_\theta(\mathbf{x})),
\label{eq:VAE}
\end{equation}

During sampling, parameter proposals are generated through the DE-MCMC mechanism \eqref{eq:DE-MCMC}. When generating the $n$-th state of the chain $\boldsymbol{\lambda}^{(n)}$, a latent sample $\mathbf{h}' \sim q_\theta(\mathbf{h} \mid \mathbf{x})$ is drawn and used to evaluate $\alpha(\boldsymbol{\lambda}^{(n-1)}, \boldsymbol{\lambda}')$ for the proposed parameter $\boldsymbol{\lambda}'$. The likelihood density $p(\mathbf{h}' \mid \boldsymbol{\lambda}')$ is approximated by the surrogate density $p_\xi(\mathbf{h}' \mid \boldsymbol{\lambda}')$ modeled via the CNF, as indicated by subscript $\xi$ and detailed in the next section. The acceptance ratio in \eq\eqref{eq:acceptance} is therefore rewritten as:
\begin{equation}
\alpha = \min\left(1,\ 
\frac{p_\xi(\mathbf{h}' \mid \boldsymbol{\lambda}') \, p(\boldsymbol{\lambda}')}
{p_\xi(\mathbf{h}' \mid \boldsymbol{\lambda}^{(n-1)}) \, p(\boldsymbol{\lambda}^{(n-1)})}
\right),
\end{equation}
which retains the Metropolis-Hastings form under a symmetric proposal. Accepted samples are used to update the Markov chain and, at regular intervals of $k_T$ iterations, to enrich the past-states archive $\mathbf{\Lambda}$. After discarding the burn-in samples and thinning the chain, the resulting sequence of samples provides an approximation to the posterior $p(\boldsymbol{\lambda} \mid \mathbf{x})$. A complete algorithmic description of the inference procedure is provided in Algorithm~\ref{alg:inference}. 

\begin{algorithm}[!h]
\hspace*{\algorithmicindent} \textbf{Input}:
Trained variational encoder $q_\theta(\mathbf{h} \mid \mathbf{x}) = \mathcal{N}(\boldsymbol{\mu}_\theta(\mathbf{x}),\, \boldsymbol{\sigma}^2_\theta(\mathbf{x}))$\\
\hspace*{49.95pt} Trained conditional normalizing flow $p_\xi(\mathbf{h} \mid \boldsymbol{\lambda})$\\
\hspace*{49.95pt} Prior distribution over target parameters $p(\boldsymbol{\lambda})$\\
\hspace*{49.95pt} Observational data $\mathbf{x}$\\
\hspace*{49.95pt} Scaling factor $\gamma$\\
\hspace*{49.95pt} Past-states archive update period $k_T$\\
\hspace*{49.95pt} Chain length $N_\text{s}$
\begin{algorithmic}[1]
\Statex
\Statex$\triangleright$ Probabilistic feature extraction
\State Compute the latent distribution $q_\theta(\mathbf{h} \mid \mathbf{x}) = \mathcal{N}(\boldsymbol{\mu}_\theta(\mathbf{x}),\,\boldsymbol{\sigma}^2_\theta(\mathbf{x}))$
\Statex
\Statex$\triangleright$ Sampling scheme
\State Initialize iteration counter $n=1$
\State Initialize parameters $\boldsymbol{\lambda}^{(1)} \sim p(\boldsymbol{\lambda})$
\State Initialize past-states archive $\mathbf{\Lambda}$
\While{$n < N_\text{s}$}
    \State $n = n+1$
    \State Sample latent variable $\mathbf{h}' \sim q_\theta(\mathbf{h} \mid \mathbf{x})$
    \State Uniformly sample past states $\mathbf{\Lambda}_{R1}, \mathbf{\Lambda}_{R2}$ from $\mathbf{\Lambda}$
    \State Evaluate candidate parameter $\boldsymbol{\lambda}'$: 
    \begin{equation*}
\boldsymbol{\lambda}' = \boldsymbol{\lambda}^{(n-1)} + \gamma (\mathbf{\Lambda}_{R1} - \mathbf{\Lambda}_{R2}) + \boldsymbol{\epsilon}
    \end{equation*}
    \State Estimate conditional likelihood density $p(\mathbf{h}' \mid \boldsymbol{\lambda}')\approx p_\xi(\mathbf{h}' \mid \boldsymbol{\lambda}')$
    \State Compute acceptance ratio:
    \begin{equation*}
\alpha = \min\left(1,\ 
\frac{p_\xi(\mathbf{h}' \mid \boldsymbol{\lambda}') \, p(\boldsymbol{\lambda}')}
{p_\xi(\mathbf{h}' \mid \boldsymbol{\lambda}^{(n-1)}) \, p(\boldsymbol{\lambda}^{(n-1)})}
\right)
    \end{equation*}
    \If{$\boldsymbol{\lambda}'$ is accepted with probability $\alpha$}
        \State $\boldsymbol{\lambda}^{(n)} \gets \boldsymbol{\lambda}'$
    \Else
        \State $\boldsymbol{\lambda}^{(n)} \gets \boldsymbol{\lambda}^{(n-1)}$
    \EndIf
\If{$n \equiv 0 \text{ }(\text{mod } k_T)$} 
 \State $\mathbf{\Lambda}=[\mathbf{\Lambda},\boldsymbol{\lambda}^{(n)}]$
 \EndIf
\EndWhile
\Statex
\Statex$\triangleright$ Post-processing (discard burn-in period and chain thinning)
\end{algorithmic}
\hspace*{\algorithmicindent} \textbf{Output}: Markov chain of samples from $p(\boldsymbol{\lambda} \mid \mathbf{h})$
\caption{Simulation-based inference of parameter posterior via \textit{neural} MCMC.}
\label{alg:inference}
\end{algorithm}

\subsection{Likelihood surrogate modeling using normalizing flows}
\label{sec:NF}
In our framework, the likelihood approximation $p(\mathbf{h}' \mid \boldsymbol{\lambda}')\approx p_\xi(\mathbf{h}' \mid \boldsymbol{\lambda}')$ exploits a neural surrogate based on a class of invertible generative models known as {\em normalizing flows}. Normalizing flows are capable of yielding highly expressive yet tractable probability distributions. In essence, they apply a sequence of invertible and differentiable transformations to map a simple base distribution, usually a standard multivariate Gaussian, into a more complex target distribution~\cite{9089305}. This construction supports both efficient sampling and effective density evaluation, and has demonstrated remarkable success across a wide range of domains, from image and audio synthesis~\cite{kingma2018glow, prenger2019waveglow} to computational statistics and simulation-based Bayesian inference~\cite{papamakarios2021normalizing}.

Formally, let $p_\mathbf{H}(\mathbf{h})$ denote a (potentially complex) probability density distribution for $\mathbf{h}$, as indicated by the subscript $\mathbf{H}$. In our setting, this density corresponds to the variational approximation $q_\theta(\mathbf{h}\mid\mathbf{x})$ produced by the VAE encoder. Let $f: \mathbb{R}^{N_\text{h}} \rightarrow \mathbb{R}^{N_\text{z}}$, with $N_\text{h}=N_\text{z}$, be an invertible transformation that maps an $\mathbf{h}$ sample to a representation $\mathbf{z}=f(\mathbf{h})$, where $\mathbf{z}\in \mathbb{R}^{N_\text{z}}$ is drawn from a base distribution $p_\mathbf{Z}(\mathbf{z})$, as indicated by the subscript $\mathbf{Z}$. The inverse mapping is given by $\mathbf{h} = g(\mathbf{z})= f^{-1}(\mathbf{z})$. By the change-of-variables formula, the density $p_\mathbf{H}(\mathbf{h})$ can be expressed as: 
\begin{equation} 
p_\mathbf{H}(\mathbf{h})=p_\mathbf{Z}(f(\mathbf{h}))\left|\det[\mathbf{D}_f(\mathbf{h})]\right|,
\label{eq:NF}
\end{equation}
where $\mathbf{D}_f(\mathbf{h})=\frac{\partial f}{\partial \mathbf{h}}\in\mathbb{R}^{N_\text{z}\times N_\text{h}}$ is the Jacobian matrix of $f$, and $\det[\mathbf{D}_f(\mathbf{h})]$ denotes its determinant. 

Let the forward transformation $f$ be constructed as a composition of $N_f$ invertible and differentiable bijections (i.e., flows) $f = f_{N_f} \circ f_{N_f-1} \circ \ldots \circ f_2 \circ f_1$. We introduce a sequence of intermediate variables $\lbrace\mathbf{h}_k\rbrace_{k=0}^{N_f}$, defined recursively as $\mathbf{h}_k=f_k(\mathbf{h}_{k-1})$, $f_k : \mathbb{R}^{N_h} \to \mathbb{R}^{N_h}$, for $k=1,\ldots,N_f$, with $\mathbf{h}_0 = \mathbf{h}$ and $\mathbf{h}_{N_f} = \mathbf{z}$. Since each $f_k$ is bijective, the overall transformation $f$ is bijective as well. Denoting $g_k = f_k^{-1}$, the inverse transformation reads $g= g_1 \circ g_2 \circ \ldots \circ g_{N_f-1} \circ g_{N_f}$. By the chain rule, the determinant of the full Jacobian $\mathbf{D}_f(\mathbf{h})$ can be written as the product of the determinants of the Jacobians of the individual sub-transformations:
\begin{equation}
    \det[\mathbf{D}_f(\mathbf{h})] = \prod_{k=1}^{N_f} \det \mathbf{D}_{f_k}(\mathbf{h}_{k-1}).
\end{equation}

Computing Jacobians and their determinants in high-dimensional spaces is generally computationally expensive. This challenge, along with the requirement for each transformation to be strictly bijective, makes the direct application of \eq\eqref{eq:NF} impractical for modeling arbitrary distributions. To address this issue, Dinh et al.~\cite{dinh2015nice} introduced a coupling-layer-based approach that enables both tractable and expressive normalizing flows. Their key insight is that the determinant of a triangular matrix can be efficiently computed as the product of its diagonal entries. This principle underlies the design of coupling layers, which are simple bijective transformations that update only a subset of the input variables while keeping the remaining components fixed. 

To implement each bijection $f_k$ as a coupling layer, we partition the input \mbox{$\mathbf{h}_{k-1}$} into two subsets $\langle\mathbf{h}^A_{k-1}, \mathbf{h}^B_{k-1}\rangle \in \mathbb{R}^{N_d}\times\mathbb{R}^{N_h-N_d}$. We then define parametrized bijections \mbox{$\widehat{f_k}(\cdot; \mathbf{c}_k) : \mathbb{R}^{N_d} \to \mathbb{R}^{N_d}$}, whose parameters \mbox{$\mathbf{c}_k=C_k(\mathbf{h}^B_{k-1};\boldsymbol{\xi}_k)$} are given by a learnable conditioning function (or conditioner) $C_k(\cdot;\boldsymbol{\xi}_k)$, with trainable weights $\boldsymbol{\xi}_k$. The resulting coupling transformation $f_k$ is defined as:
\begin{equation}
    \mathbf{h}_k  =
    \begin{pmatrix} \mathbf{h}^A_k\\ \mathbf{h}^B_k\end{pmatrix} = 
    f_k(\mathbf{h}_{k-1})=\begin{pmatrix}
    \widehat{f}_k(\mathbf{h}^A_{k-1}; C_k(\mathbf{h}^B_{k-1};\boldsymbol{\xi}_k))\\
    \mathbf{h}^B_{k-1} \end{pmatrix},
\end{equation}
where the conditioner $C_k$ uses only the masked input $\mathbf{h}^B_{k-1}$ to parametrize the transformation $\widehat{f}_k$ applied to $\mathbf{h}^A_{k-1}$. By construction, $f_k$ is invertible provided that $\widehat{f}_k$ is invertible. Moreover, due to the input partitioning, the Jacobian matrix $\mathbf{D}_{f_k}$ is block-triangular:
\begin{equation}
    \mathbf{D}_{f_k}(\mathbf{h}_{k-1}) =
\begin{bmatrix}
\mathbf{D}_{\widehat{f}_k}(\mathbf{h}^A_{k-1}) & \frac{\partial \mathbf{h}^A_{k}}{\partial \mathbf{h}^B_{k-1}} \\
     0 & \mathbf{I}_{N_h-N_d}
\end{bmatrix},
\label{eq:block}
\end{equation}
where $\mathbf{I}_{N_h-N_d}$ denotes the identity matrix of size $N_h-N_d$. As a result, the determinant of $\mathbf{D}_{f_k}(\mathbf{h}_{k-1})$ reduces to the determinant of $\mathbf{D}_{\widehat{f}_k}(\mathbf{h}^A_{k-1})$, which can be computed efficiently.

\begin{figure}[t]
    \centering
\includegraphics[width=0.98\linewidth]{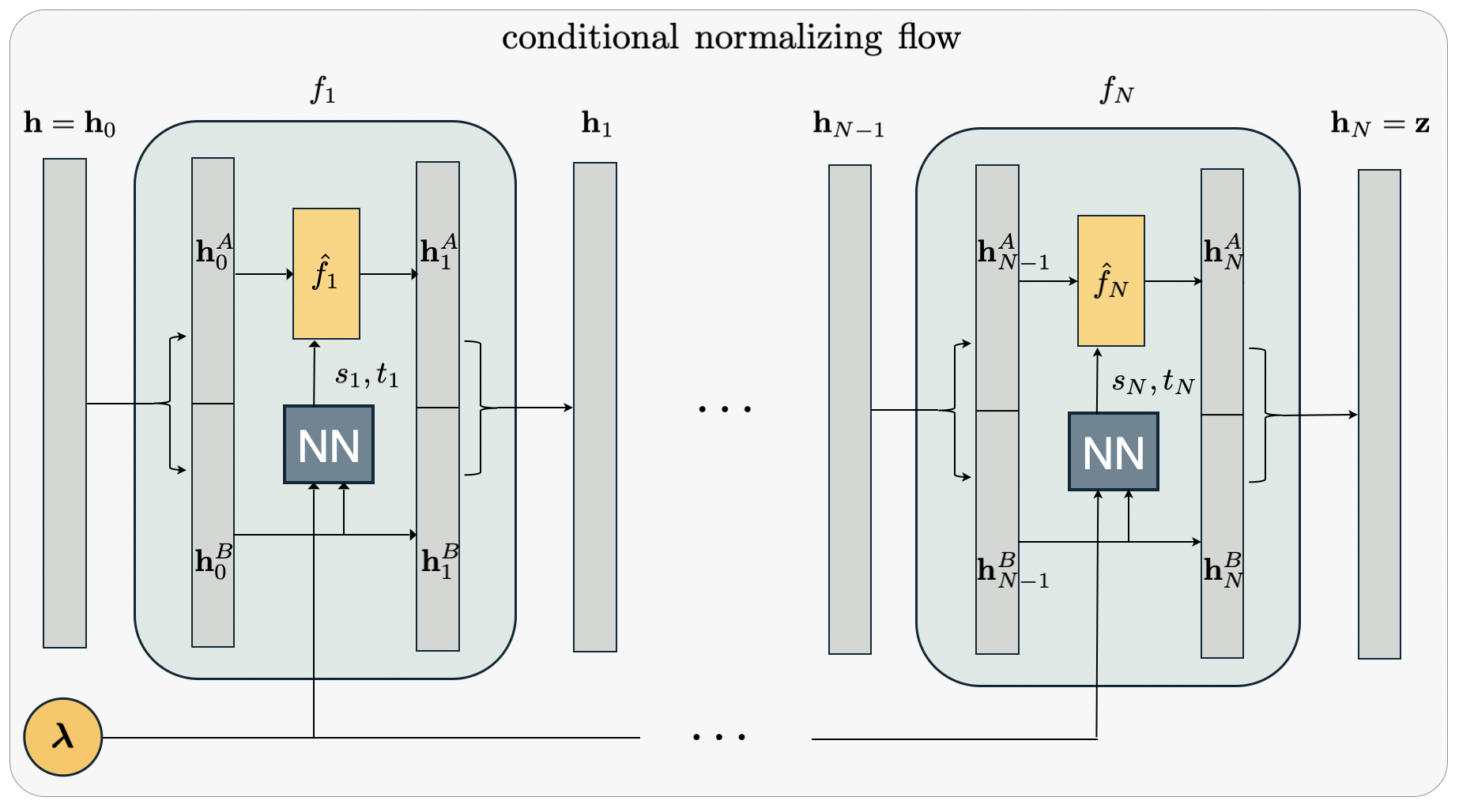}
     
    \caption{Schematic of the real-valued non-volume preserving normalizing flow with the additional conditioning structure on $\boldsymbol{\lambda}$ for estimating the likelihood density $p(\mathbf{h} \mid \boldsymbol{\lambda})$. Following the normalizing direction $\mathbf{z}=f(\mathbf{h})$, a sample $\mathbf{h}$ from the variational approximation $q_\theta(\mathbf{h}\mid\mathbf{x})$ is transformed into a realization $\mathbf{z}$ from the base distribution through a sequence of conditional affine coupling layers $f=f_{N} \circ f_{N-1} \circ \ldots \circ f_2 \circ f_1$ conditioned on $\boldsymbol{\lambda}$. For notational simplicity, the total number of flow layers is denoted by $N$ in the figure, rather than $N_f$.}
    \label{fig:flow}
\end{figure}

In this work, we adopt affine coupling layers, as used in RealNVP architectures~\cite{dinh2017realnvp}, which are particularly effective for density estimation in high-dimensional settings. The transformation $\widehat{f}_k$ applied to $\mathbf{h}^A_{k-1}$ therefore consists of an element-wise scaling and shifting:
\begin{equation}
    \mathbf{h}^A_k =  \widehat{f}_k(\mathbf{h}^A_{k-1}; C_k(\mathbf{h}^B_{k-1};\boldsymbol{\xi}_k)) = \mathbf{h}^A_{k-1} \odot \exp(s_k(\mathbf{h}^B_{k-1})) + t_k(\mathbf{h}^B_{k-1}), \label{eq:AffineCL1} 
\end{equation}
where $s_k:\mathbb{R}^{N_h-N_d}\mapsto\mathbb{R}^{N_d}$ and $t_k:\mathbb{R}^{N_h-N_d}\mapsto\mathbb{R}^{N_d}$ denote the scale and translation functions, respectively, and $\odot:(\mathbb{R},\mathbb{R})\mapsto\mathbb{R}$ denotes element-wise multiplication. Both $s_k$ and $t_k$ are produced by the conditioner $C_k$, which embeds the masked input $\mathbf{h}^B_{k-1}$ into $\widehat{f}_k$. The corresponding Jacobian $\mathbf{D}_{\widehat{f}_k}(\mathbf{h}^A_{k-1})$ is diagonal, with entries $\text{diag}\left[\exp(s_k(\mathbf{h}^B_{k-1}))\right]$. Consequently, the determinant of the full Jacobian $\mathbf{D}_{f_k}(\mathbf{h}_{k-1})$ simplifies to
\begin{equation}
     \det \mathbf{D}_{f_k}(\mathbf{h}_{k-1}) = \exp \left( \sum_{m=1}^{N_d} s_k(\mathbf{h}^B_{k-1})_{m} \right),
\end{equation}
which can be computed efficiently without requiring the Jacobians of $s_k$ or $t_k$. This feature allows $s_k$ and $t_k$ to be arbitrarily complex functions. Moreover, the inverse transformation can be computed as efficiently as the forward pass. Indeed, inverting the affine coupling layer does not require computing the inverses of $s_k$ or $t_k$, thereby enabling the use of expressive, potentially non-invertible neural networks. Furthermore, as schematized in \fig\ref{fig:flow}, we condition the computation of $s_k$ and $t_k$ on label information $\boldsymbol{\lambda}$, following the approach proposed in~\cite{xiao2019method}. More formally, the coupling transformation $\widehat{f}_k$ is parametrized as $\widehat{f}_k(\mathbf{h}^A_{k-1}; C_k(\mathbf{h}^B_{k-1}, \boldsymbol{\lambda};\xi_k))$, where the learnable conditioning network $C_k$ takes both the masked features $\mathbf{h}^B_{k-1}$ and label information $\boldsymbol{\lambda}$ as inputs. The scale and translation functions are thus defined as $s_k=s_k(\mathbf{h}^B_{k-1}, \boldsymbol{\lambda})$ and $t_k=t_k(\mathbf{h}^B_{k-1}, \boldsymbol{\lambda})$,  respectively. This conditional affine coupling layer architecture enables efficient and accurate modeling of our \textit{neural} MCMC likelihood mechanism.

Let $\mathcal{D}_\mathbf{h}=\lbrace(\mathbf{h}_i,\boldsymbol{\lambda}_i)\rbrace_{i=1}^{N_I}$ denote a dataset of $N_I$ feature-label instances, where each feature $\mathbf{h}_i$ (derived from raw observations $\mathbf{x}_i$) is associated with a target parameter vector $\boldsymbol{\lambda}_i$, $i=1,\ldots,N_I$. The CNF is trained to assign higher likelihoods to correctly labeled samples and lower likelihoods to incorrect ones, implicitly discriminating true or near-true label assignments from incorrect ones. Specifically, the set of learnable weights $\mathbf{\Theta}_\text{CNF}=\lbrace\boldsymbol{\xi}_1,\ldots,\boldsymbol{\xi}_{N_f}\rbrace$ parametrizing the CFN is optimized by minimizing the negative log-likelihood of the observed data under the model:
\begin{align}
    \mathbf{\Theta}_\text{CNF}^*&= \argmin_{\mathbf{\Theta}_\text{CNF}} \mathcal{L}_{\text{CNF}}(\mathcal{D}_\mathbf{h}, \mathbf{\Theta}_\text{CNF}),\\
    \mathcal{L}_{\text{CNF}}&= - \frac{1}{N_I} \sum_{i=1}^{N_I}\log p_\xi(\mathbf{h}_i\mid\boldsymbol{\lambda}_i)= - \frac{1}{N_I} \sum_{i=1}^{N_I}\biggl[\log p_\mathbf{Z}(f(\mathbf{h}_i;\boldsymbol{\lambda}_i)) + \log\left|\det[\mathbf{D}_f(\mathbf{h}_i;\boldsymbol{\lambda}_i)]\right|\biggr].
\label{eq:NLL}
\end{align}
which explicitly depends on the conditioning $\boldsymbol{\lambda}_i$. The objective in \eq\eqref{eq:NLL} corresponds to the empirical negative conditional log-likelihood, obtained via the change-of-variables formula \eqref{eq:NF}, where $f( \cdot;\boldsymbol{\lambda})$ denotes the invertible transformation induced by the conditional flow and $\mathbf{D}_f(\cdot;\boldsymbol{\lambda})$ its Jacobian. It is worth noting that the logarithm formulation ensures numerical stability during gradient-based optimization. 

This architecture offers a significant advantage over standard feed-forward neural networks as it enables exact likelihood evaluation through the change-of-variables formula. This capability makes conditional normalizing flows particularly well-suited for Bayesian inference, where accurate density evaluation is essential. In contrast, conventional neural networks typically learn deterministic mappings without defining consistent probability density functions over their output space. This difference is also reflected in the training process. Normalizing flows are trained by directly minimizing the negative log-likelihood of the data, as shown in \eqref{eq:NLL}, thereby preserving a clear probabilistic interpretation without relying on surrogate objectives. Minimizing this loss encourages the learned transformation to map samples from the latent distribution $q_\theta(\mathbf{h}\mid\mathbf{x})$ to the base distribution $p_\mathbf{Z}(\mathbf{z})$, conditioned on $\boldsymbol{\lambda}$, while maintaining exact likelihood evaluation. By contrast, standard DL approaches to density estimation often rely on indirect or approximate training objectives. Such approximations can struggle to capture complex dependencies or to represent multimodal, skewed, or highly correlated distributions.

\subsection{Variational autoencoder for probabilistic feature extraction}
\label{sec:VAE}

Applying a normalizing flow directly to high-dimensional observational data can be computationally inefficient and prone to overfitting. Moreover, observation noise can obscure the underlying signal structure, degrading the quality of the learned likelihood model. For these reasons, we preliminary compress data into a lower-dimensional latent space using a variational autoencoder.

Autoencoders are DL architectures designed for self-supervised representation learning. An AE consists of two main components: an encoder and a decoder, typically connected through a central bottleneck layer whose dimensionality is smaller than that of both the input and output layers. The model learns to encode the input data into a compact latent space and then decode it to reconstruct the original input. As a result, the learned latent representation must preserve the essential information required for reconstruction, enabling feature extraction capabilities beyond traditional linear projection techniques.

In this work, we employ VAE-based feature extraction, which extends standard autoencoders by replacing the deterministic latent representation with a probabilistic one. Instead of mapping an input $\mathbf{x}$ to a single point in the latent space, the encoder outputs the parameters of a distribution over latent variables. Assuming the Gaussian variational posterior $q_\theta(\mathbf{h} \mid \mathbf{x})$, the encoder defines the probabilistic encoding in \eq\eqref{eq:VAE}. To enable stochastic sampling from $q_\theta(\mathbf{h} \mid \mathbf{x})$ while preserving differentiability during training, we adopt the reparameterization trick~\cite{kingma2013auto}. Specifically, latent samples are expressed as: 
\begin{equation}
\mathbf{h}=\boldsymbol{\mu}_\theta(\mathbf{x}) + \boldsymbol{\sigma}_\theta(\mathbf{x})\odot\boldsymbol{\epsilon_\text{VAE}},\qquad \boldsymbol{\epsilon_\text{VAE}}\sim\mathcal{N}(\mathbf{0}_{N_h},\mathbf{I}_{N_h}),
\end{equation}
which allows gradients to propagate through the sampling operation.  Given a latent realization $\mathbf{h}$, the decoder $d_\theta:\mathbb{R}^{N_h}\mapsto\mathbb{R}^{N_x}$ is used to reconstruct the original signal in the observation space:  
\begin{equation}
   \widehat{\mathbf{x}}=d_\theta(\mathbf{h}),    
   \label{eq:dec}
\end{equation}
where $\widehat{\mathbf{x}}\in\mathbb{R}^{N_x}$ denotes the reconstructed input.

The encoder and decoder are jointly trained to maximize a variational lower bound on the marginal log-likelihood of the observed data, known as the evidence lower bound (ELBO). This objective provides a tractable surrogate for the generally intractable data likelihood. By balancing reconstruction fidelity with latent-space regularization, ELBO maximization yields a smooth and well-structured latent manifold suitable for generative modeling and downstream inference tasks. For a single observation $\mathbf{x}$, the ELBO is given by:
\begin{equation}
\log p(\mathbf{x}) \;\ge\;
\mathbb{E}_{q_{\theta}(\mathbf{h}\mid \mathbf{x})}
\!\left[ \log p_\theta(\mathbf{x}\mid \mathbf{h}) \right]
\;-\;
\mathrm{D}_{\mathrm{KL}}\!\left[
q_\theta(\mathbf{h}\mid \mathbf{x})
\,\middle\|\,
\overline{p}(\mathbf{h})
\right],
\label{eq:ELBO}
\end{equation}
where $\mathbb{E}_{q_{\theta}(\mathbf{h}\mid \mathbf{x})}[\ \cdot\ ]$ denotes the expectation with respect to the variational posterior, and $\mathrm{D}_{\mathrm{KL}}\left[\ \cdot \mid\mid \cdot \ \right]$ denotes the Kullback--Leibler (KL) divergence between two probability distributions~\cite{kullback1951information}. The conditional Gaussian likelihood $p_\theta(\mathbf{x}\mid \mathbf{h})=\mathcal{N}(\mathbf{x}; d_\theta(\mathbf{h}),\sigma_x^2\mathbf{I}_{N_x})$ is induced by the decoder and assumes a mean given by the deterministic decoding function $d_\theta(\mathbf{h})$ and fixed diagonal covariance. The observation noise level $\sigma_x\in\mathbb{R}$ is distinct from the standard deviation  $\boldsymbol{\sigma}_\theta$ of the variational posterior and is typically fixed (e.g., $\sigma_x=1$). Finally, $\overline{p}(\mathbf{h})$ denotes a prescribed prior over the latent variables, commonly chosen as a standard normal distribution. 

The first term in \eq\eqref{eq:ELBO} corresponds to a reconstruction objective. Under the assumption of a Gaussian likelihood, it reduces, up to additive constants independent of the model parameters, to a mean squared error (MSE) loss. Let $\mathcal{D}_\mathbf{x}=\lbrace \mathbf{x}_i\rbrace_{i=1}^{N_I}$ denote a dataset of $N_I$ observation instances. Denoting the set of trainable parameters of the VAE by $\mathbf{\Theta}_\text{VAE}$, the reconstruction loss is defined as
\begin{equation}
\mathcal{L}_{\text{MSE}}(\mathcal{D}_\mathbf{x}, \mathbf{\Theta}_\text{VAE})
    = \frac{1}{N_I} \sum_{i=1}^{N_I} 
    -\frac{1}{2\sigma_x^2}\lVert\mathbf{x}_i-d_\theta(\mathbf{h}_i)\rVert_2^2,
\end{equation}
which simplifies optimization while retaining a probabilistic interpretation.

The second term in \eq\eqref{eq:ELBO} acts as a regularizer that constrains the learned latent distribution to remain close to the prior. For Gaussian variational posteriors, the KL divergence admits the followin closed-form
expression:
\begin{equation}
    \mathrm{D}_\mathrm{KL}\!\left[
        q_\theta(\mathbf{h}\mid\mathbf{x}_i)
        \,\middle\|\,
        \overline{p}(\mathbf{h})
    \right]
    = \frac{1}{2} \sum_{j=1}^{N_h}
      \left(
        1 + \log \sigma_j^2(\mathbf{x}_i)
        - \mu_j^2(\mathbf{x}_i)
        - \sigma_j^2(\mathbf{x}_i)
      \right),
      \label{eq:kl}
\end{equation}
where $\mu_j(\mathbf{x}_i)$ and $\sigma_j^2(\mathbf{x}_i)$ denote the mean and
variance of the $j$-th latent dimension, $j=1,\ldots,N_h$. It is worth noting that this expression is equivalent to the standard KL divergence up to a sign convention, consistent with loss minimization. The corresponding loss component is defined as:
\begin{equation}
    \mathcal{L}_\mathrm{KL}(\mathcal{D}_{\mathbf{x}}, \mathbf{\Theta}_\text{VAE})
    = \frac{1}{N_I}
      \sum_{i=1}^{N_I}
      \mathrm{D}_\mathrm{KL}\!\left[
        q_\theta(\mathbf{h}\mid\mathbf{x}_i)
        \,\middle\|\, 
        \overline{p}(\mathbf{h})
      \right].
    \label{eq:KL}
\end{equation}

Combining reconstruction and regularization terms, the VAE parameters are learned by solving the following optimization problem:
\begin{equation}
    \mathbf{\Theta}_\mathrm{VAE}^*
    = \argmin_{\mathbf{\Theta}_\mathrm{VAE}}
      \left[
        \mathcal{L}_\mathrm{MSE}(\mathcal{D}_{\mathbf{x}}, \mathbf{\Theta}_\mathrm{VAE})
        + 
        \beta\,
        \mathcal{L}_\mathrm{KL}(\mathcal{D}_{\mathbf{x}}, \mathbf{\theta}_\mathrm{VAE})
      \right],
\end{equation}
where $\beta_{\text{KL}}\in\mathbb{R}^+$ controls the strength of the latent-space regularization.

Forward models are inevitably affected by model-form and epistemic uncertainties. Despite efforts to refine and calibrate them, the inherent complexity of physical and engineering systems often imposes limitations on modeling fidelity. A key concern that arises in this context is the risk of producing overconfident yet inaccurate estimates. Latent variable models such as VAEs can help bridge the gaps left by deterministic feature extraction techniques. In particular, their probabilistic structure can be leveraged to represent discrepancies between the true data-generating process and the distribution learned from training data, thereby enabling more robust and uncertainty-aware feature extraction.

\begin{figure}[t]
    \centering
\includegraphics[width=0.98\linewidth]{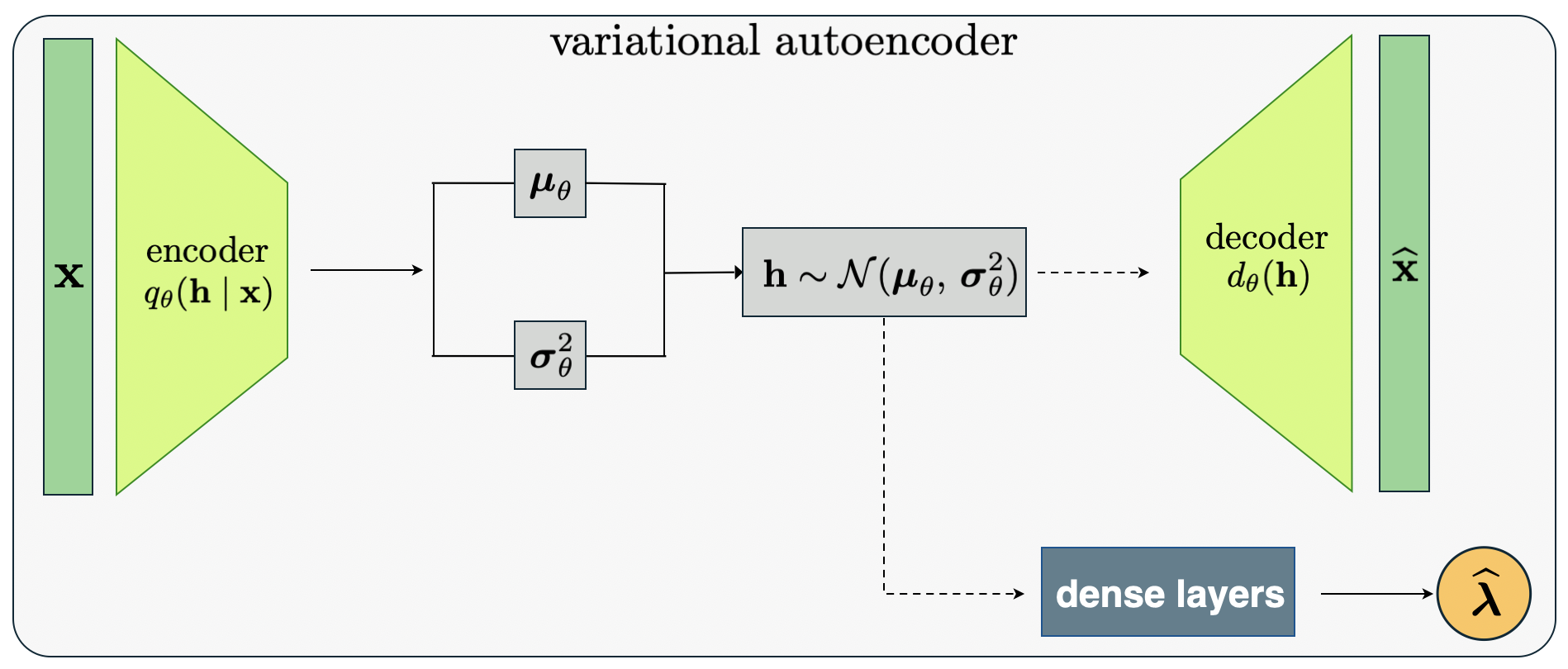}
    \caption{Schematic of the supervised variational autoencoder with the additional prediction block, enabling a label-informed representation for task-specific inference. Hat variables indicate neural network approximations.}
    \label{fig:VAE}
\end{figure}

While standard VAEs are self-supervised, our implementation augments this architecture by introducing a dense prediction head branching from the bottleneck layer, as shown in \fig\ref{fig:VAE}. We leverage this hybrid unsupervised-supervised formulation to learn a regularized latent space that is also enriched with label-relevant information, making the resulting representation more suitable for downstream inference of target variables. 

During training, samples drawn from the latent posterior distribution are propagated through two parallel pathways: (i) one feeds the decoder for input reconstruction, and (ii) the other routes the latent samples through a stack of dense layers to produce task-specific predictions. This latter prediction branch introduces an additional supervised loss term into the overall training objective. In our applications, the labels correspond to  the vectors of target parameters $\boldsymbol{\lambda}$. Accordingly, the prediction task is formulated as a regression problem, and the associated loss is defined as the MSE between the true and predicted parameter vectors:
\begin{equation}
\mathcal{L}_{\text{pred}} = \frac{1}{N_I} \sum_{i=1}^{N_I} \lVert \boldsymbol{\lambda}_i - \widehat{\boldsymbol{\lambda}}_i\rVert_2^2,
\end{equation}
where $\widehat{\boldsymbol{\lambda}}_i$ denotes the predicted parameter vector associated with the $i$-th input sample.

Let $\mathcal{D}_{\boldsymbol{\lambda}}=\lbrace(\mathbf{x}_i,\boldsymbol{\lambda}_i)\rbrace_{i=1}^{N_I}$ denote the dataset of input-label pairs, where each input $\mathbf{x}_i$ is associated with a corresponding target parameter vector $\boldsymbol{\lambda}_i$. Denoting the set of learnable parameters of the informed-VAE (i-VAE) architecture by $\mathbf{\Theta}_\text{i-VAE}$, the learning problem is formulated as the following optimization: 
\begin{equation}
    \mathbf{\Theta}_\text{i-VAE}^* = \argmin_{\mathbf{\Theta}_\text{i-VAE}} \mathcal{L}_{\text{i-VAE}}(\mathcal{D}_{\boldsymbol{\lambda}}, \mathbf{\Theta}_\text{i-VAE}),
\end{equation}
where $\mathcal{L}_{\text{i-VAE}}$ is the total loss function, given by:
\begin{equation}
\mathcal{L}_{\text{i-VAE}} = \mathcal{L}_{\text{MSE}} + \beta_{\text{KL}}\mathcal{L}_{\text{KL}} + \beta_{\text{pred}}\mathcal{L}_{\text{pred}},
\end{equation}
where $\mathcal{L}_{\text{MSE}}$ enforces input reconstruction, $\mathcal{L}_{\text{KL}}$ regularizes the latent space, and $\mathcal{L}_{\text{pred}}$ promotes the learning of latent representations that are informative for parameter prediction. The hyperparameters $\beta_{\text{KL}},\beta_{\text{pred}}\in\mathbb{R}^+$ modulate the relative contributions of the loss terms, thereby controlling the trade-off between latent space structure, reconstruction fidelity, and inference accuracy.

\section{Numerical Experiments}
\label{sec:3}
The proposed computational methodology is demonstrated through two case studies: (i) the structural health monitoring of a railway bridge, and (ii) the identification of a spatially varying hydraulic conductivity field in a groundwater flow system. 

The full-order and reduced-order models for the first case study have been implemented in the \texttt{Matlab} environment using the \texttt{redbKIT} library~\cite{Redbkit}. The full-order problem for the second case study has been solved using the \texttt{Python} library \texttt{FEniCS}~\cite{alnaes2015fenics}. The DL architectures have been implemented using the \texttt{Tensorflow}-based \texttt{Keras} API~\cite{Keras}, while the MCMC simulations have been carried out with the open-source \texttt{Python} library \texttt{PyMC}~\cite{PyMC}. All computations have been performed on a MacBook Pro (Apple M1, 2020)
with a unified memory architecture and 16~GB of RAM.

\subsection{Railway bridge case study}
\begin{figure}[!b]
    \centering
    \includegraphics[width=0.55\textwidth]{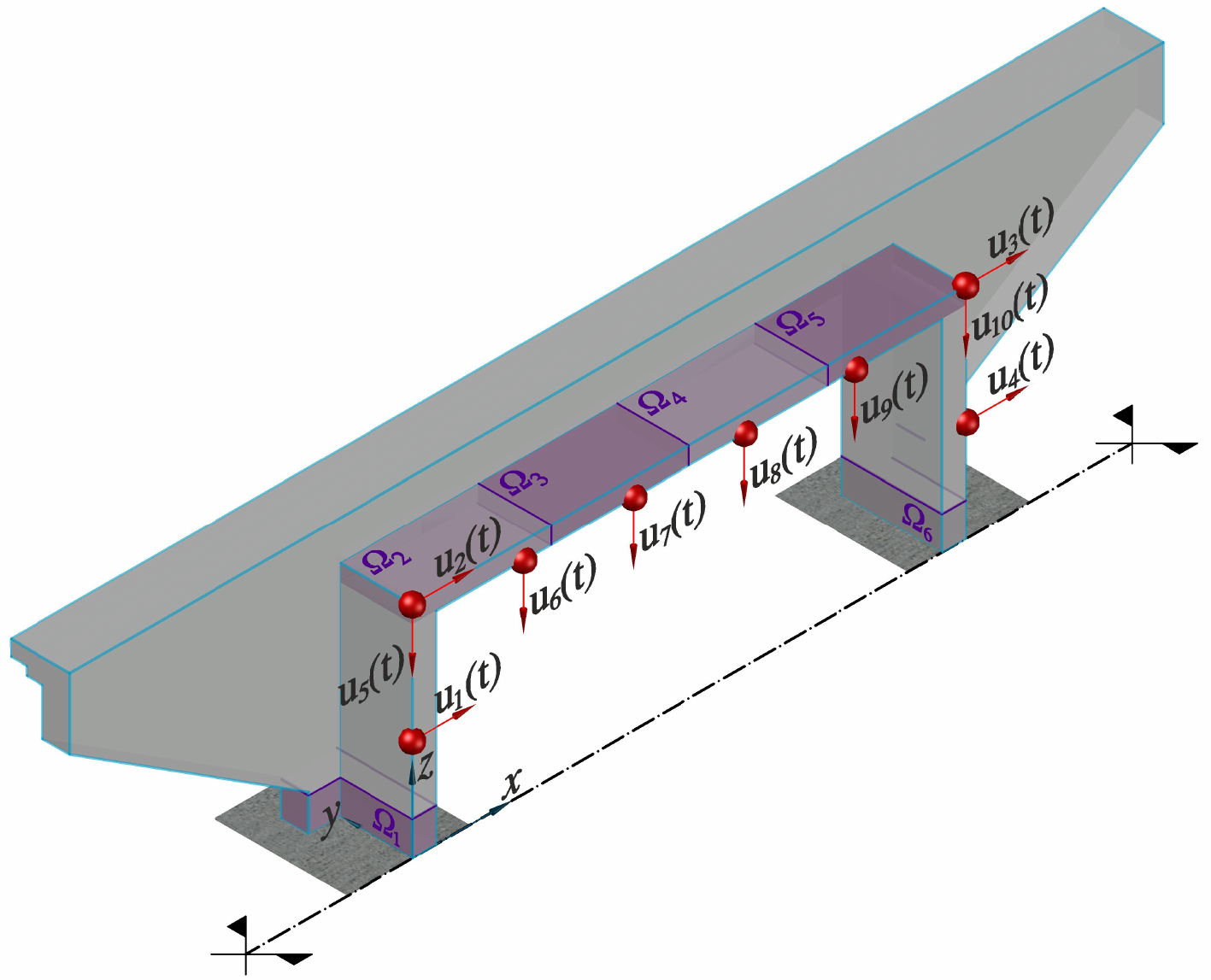}
    \caption{Railway bridge - Schematization for structural health monitoring, including details of synthetic recordings related to displacements $u_1(t),\ldots,u_{10}(t)$, and predefined damage regions $\Omega_1,\ldots,\Omega_6$.} 
    \label{fig:railwaybridge}
\end{figure}

The first case study concerns the H{\"o}rnefors railway bridge. A schematization of the bridge for structural health monitoring purposes is shown in \fig\ref{fig:railwaybridge}. The geometrical and mechanical properties, as well as the overall computational setup, are adapted from former research activities~\cite{torzoni2024digital,Torzoni_DML,5_TMM,Torzoni_MF}. 
The bridge has a span of $15.7~\text{m}$, a free height of $4.7~\text{m}$, and a width of $5.9~\text{m}$ (excluding the edge beams). The wing walls extend longitudinally up to $8~\text{m}$. The structural elements have thicknesses of $0.5~\text{m}$ for the deck, $0.7~\text{m}$ for the frame walls, and $0.8~\text{m}$ for the wing walls. The bridge is founded on two plates connected by two stay beams and supported by pile groups. The concrete is of class C35/45, with Young’s modulus $E=34~\text{GPa}$, Poisson’s ratio $\nu= 0.2$, and density $\rho=2500~\text{kg/m}^3$. The superstructure consists of a single track with sleepers spaced $0.65~\text{m}$ apart, resting on a ballast layer $0.6~\text{m}$ deep and $4.3~\text{m}$ wide, with density $\rho_B=1800~\text{kg/m}^3$. The bridge supports the transit of 8-axles \text{Gr\"{o}na T\r{a}get} trains traveling at speeds in the range $v\in[160,215]~\text{km/h}$, with each axles carrying a mass of $m_A\in[16,22]~\text{ton}$.

The data for this test case are synthetically generated using the numerical model described below. This approach allows full control over damage scenarios and operational conditions for training and validation of the proposed inference framework. The monitoring system provides displacement recordings $\mathbf{U}(\boldsymbol{\mu}) =[\mathbf{u}_1(\boldsymbol{\mu}), \ldots, \mathbf{u}_{N_u}(\boldsymbol{\mu})]\in\mathbb{R}^{L\times N_s}$, consisting of $N_s=10$ time series corresponding to degrees of freedom highlighted in \fig\ref{fig:railwaybridge}. Each time series contains $L=600$ displacement measurements uniformly sampled over the time interval $[0,T=1.5~\text{s}]$. These recordings coincide with the raw measurement vector $\mathbf{x}$ used as input to the neural MCMC framework described in \sez\ref{sec:2}. The vector $\boldsymbol{\mu} \in\mathbb{R}^{N_\mu}$ collects $N_\mu$ parameters characterizing the operational and damage conditions. For the assumed problem setting, each observation window is sufficiently short for these parameters to be regarded as time-invariant, yet long enough to enable a reliable identification of the structural behavior. To account for measurement uncertainty, the displacement recordings are contaminated with additive Gaussian noise, yielding a signal-to-noise ratio of 100.

Changes in the structural dynamic response due to damage are modeled through a localized reduction of the effective stiffness, under the assumption of a clear time-scale separation between damage evolution and health assessment. The local stiffness degradation is parameterized by two variables, $y\in\mathbb{N}$ and $\delta\in\mathbb{R}$, both included in the parameter vector $\boldsymbol{\mu}$. Specifically, $y\in\lbrace0,\ldots,6\rbrace$ identifies the damage region among a set of $N_\Omega=6$ predefined damageable subdomains $\Omega_1,\ldots,\Omega_6$, as illustrated in \fig\ref{fig:railwaybridge}, with $y=0$ denoting the undamaged reference configuration. Within a damaged region, the stiffness is reduced of a factor $\delta\in[30\%,80\%]$, kept fixed while a train travels across the bridge.  

The full-order model features $17,292$ degrees of freedom, resulting from a FE discretization using linear tetrahedral finite elements with an element size of $0.80~\textup{m}$, locally refined to $0.15~\textup{m}$ in the deck region. The ballast layer is modeled by assigning an increased density to the deck and edge beams. The embankments are represented by distributed springs, implemented as Robin boundary conditions (with elastic coefficient $10^{8}~\textup{N/m}^3$) applied to the surfaces in contact with the ground. Structural dissipation is accounted for through Rayleigh damping, calibrated to yield a $5\%$ damping ratio for the first two structural modes. The solution time interval is uniformly partitioned into $600$ time steps, and the time integration is carried out using the implicit, unconditionally stable constant-average-acceleration Newmark scheme~\cite{hughes2000finite}. 

A projection-based reduced-order model is employed to speed up the offline dataset generation phase, following the approach adopted in~\cite{Torzoni_DML}. Specifically, to reduce the computational cost associated with repeatedly solving the full-order model for different realizations of $\boldsymbol{\mu}$, a reduced-order model is constructed using a proper orthogonal decomposition (POD)-Galerkin reduced basis method~\cite{quarteroni2015reduced}. The reduced-order model is derived from a snapshot matrix assembled from $400$ full-order solutions corresponding to different parameter realizations $\boldsymbol{\mu} = (v,m_A,y,\delta)^\top$, which are taken as uniformly distributed and sampled via Latin hypercube rule. By enforcing a tolerance $10^{-3}$ on the fraction of disregarded energy, the resulting reduced basis is made by $133$ POD modes. This reduced-order model is then used to populate the training dataset $\mathcal{D}^\text{train}_{\boldsymbol{\lambda}}$ with $10,000$ instances. These data are subsequently employed to train the DL models as described in \sez\ref{sec:2}. In contrast, a testing dataset $\mathcal{D}^\text{test}_{\boldsymbol{\lambda}}$ consisting of $4000$ instances is generated using the full-order model. 

The parameter vector is organized as $\boldsymbol{\lambda}=(\delta(\Omega_1),\ldots,\delta(\Omega_6))^\top$, with $N_\Omega = 6$ entries representing the damage magnitude within the damageable subdomains $\Omega_1,\ldots,\Omega_6$, where the $i$-th component is denoted by $\lambda_i = \delta(\Omega_i)$. The low-dimensional feature space is set to dimension $N_h=4$, which has been found to provide a sufficiently expressive yet compact representation for the CNF to compute accurate log-likelihood values. 

\begin{figure}[!t]
    \centering
    \includegraphics[width=0.49\linewidth]{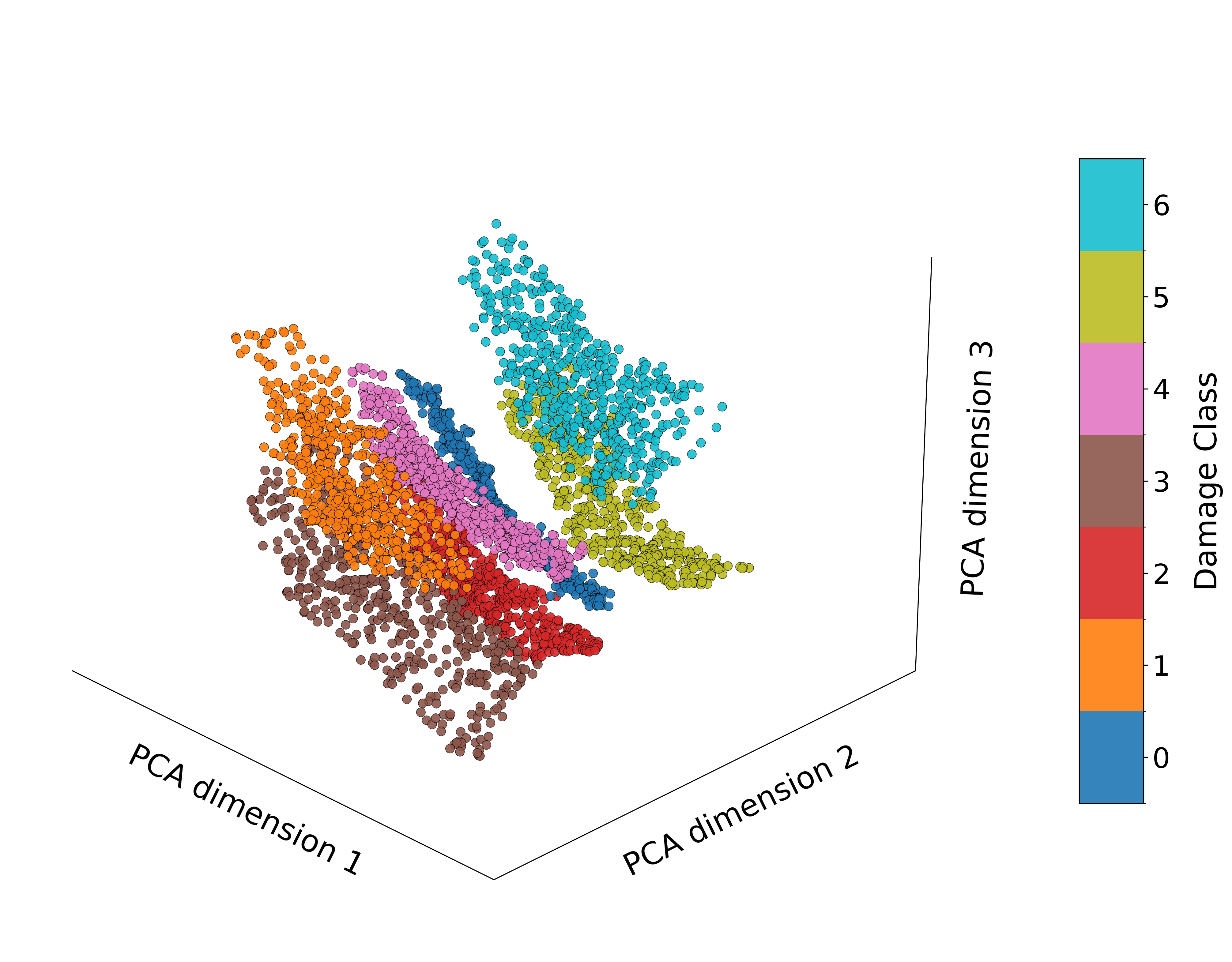}
    \includegraphics[width=0.49\linewidth]{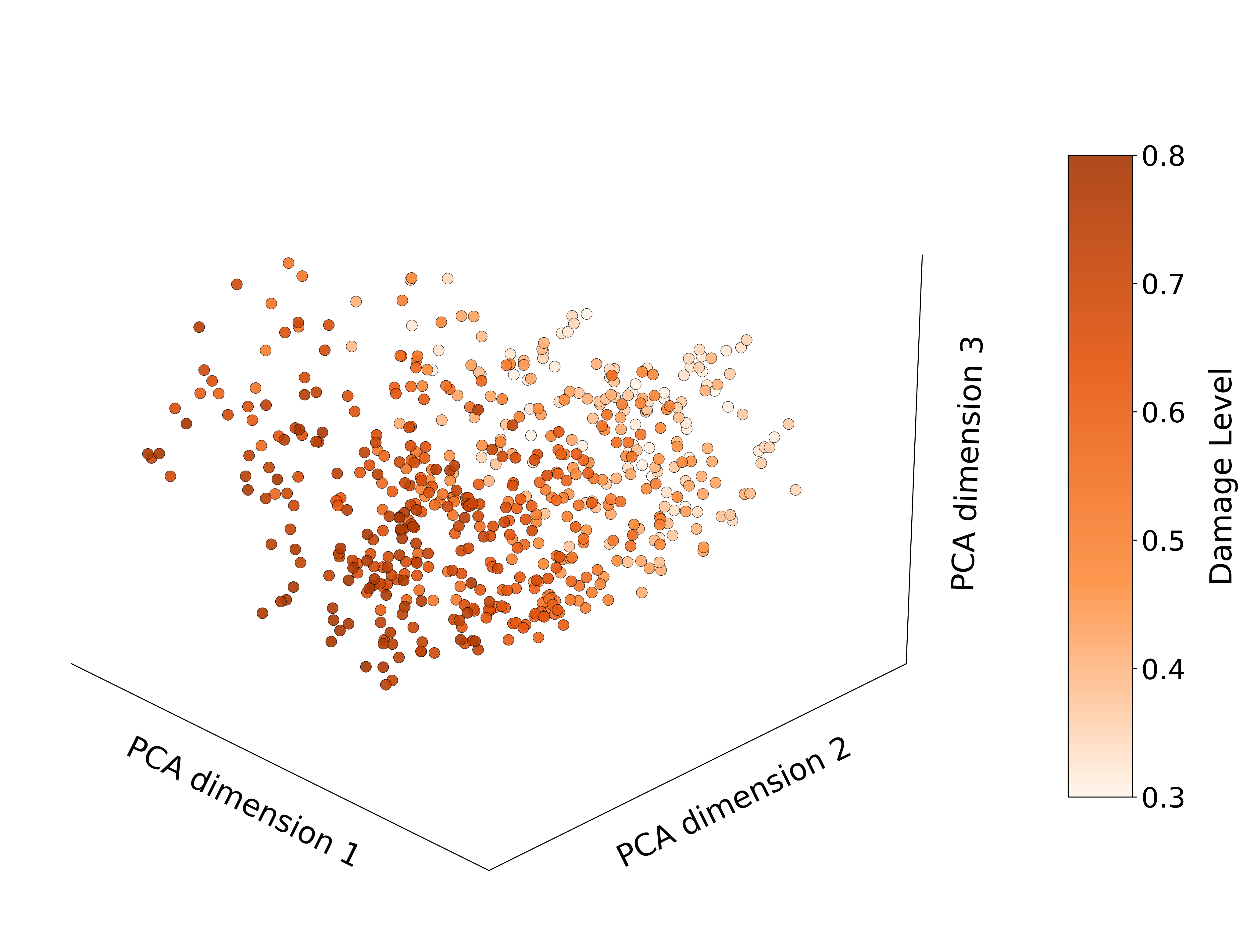}
    \caption{Railway bridge - Three-dimensional principal component  projections of the latent-space mean vectors for the testing data. Left: data colored according to the target damage location class. Right: testing data belonging to damage class $y=1$, colored according to the target damage magnitude.}
    \label{fig:pca-clust}
\end{figure}

A three-dimensional PCA projection of the $N_h$-dimensional latent mean vectors extracted by the i-VAE for each instance in the test set $\mathcal{D}^\text{test}_{\boldsymbol{\lambda}}$ is shown in \fig\ref{fig:pca-clust} (left). The color coding corresponds to the damage class $y$, revealing a clear cluster structure that reflects the sensitivity of the measured structural response to the damage location. 
Beyond damage localization, \fig\ref{fig:pca-clust} (right) presents an analogous visualization for the damage level parameter $\delta$, obtained by displaying only samples belonging to class $y=1$. These scatter plots qualitatively illustrate how the i-VAE extracts informative latent representations, thereby enabling the CNF to produce accurate log-likelihood evaluations.

\begin{figure}[!t]
    \centering
    \includegraphics[width=0.7\linewidth]{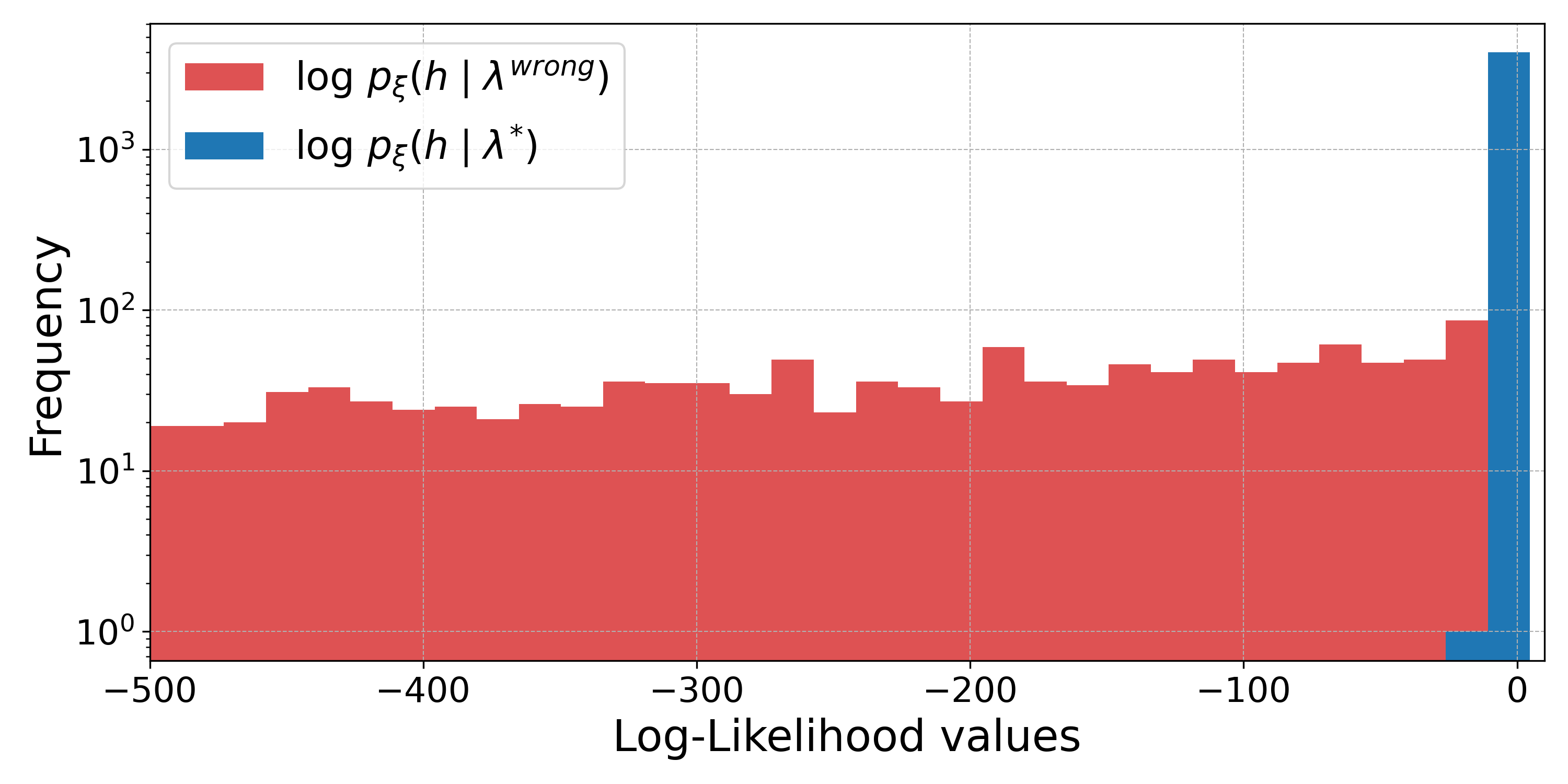}
    \caption{Railway bridge - Frequency distribution of the log-likelihood values for latent representations paired with correct labels (blue) and incorrect labels (red).}
    \label{fig:NVPbridge}
\end{figure}

To assess the capability of the CNF to assign high log-likelihood values to correctly labeled samples and lower values to incorrect ones, \fig\ref{fig:NVPbridge} reports the log-likelihoods computed for the testing data. The CNF is evaluated on each test sample twice: once paired with its true label vector $\boldsymbol{\lambda}^{*}$, and once paired with a randomly selected label vector $\boldsymbol{\lambda}^{\textit{wrong}}$ drawn from the testing dataset $\mathcal{D}^\text{test}_{\boldsymbol{\lambda}}$. Log-likelihood values associated with the correct parameter vectors are shown in blue, while those corresponding to incorrect labels are shown in red. The model consistently assigns higher log-likelihood values to the correct labels, demonstrating its surrogate likelihood model capability.

The DE-MCMC algorithm is initialized with a uniform prior $p(\boldsymbol{\lambda})$, which is iteratively updated to the target posterior distribution $p(\boldsymbol{\lambda}\mid\mathbf{U})$ by running two chains sequentially, each for $N_\text{s} = 20,000$ iterations. This number of samples has been selected to ensure a Gelman-Rubin statistic below 1.01, thereby assessing reliable convergence to a stationary distribution~\cite{art:Gelman-Rubin}. The runtime for the each chain is approximately 1.5 minutes.

\begin{figure}[!t]
    \centering
    \includegraphics[width=0.8\linewidth]{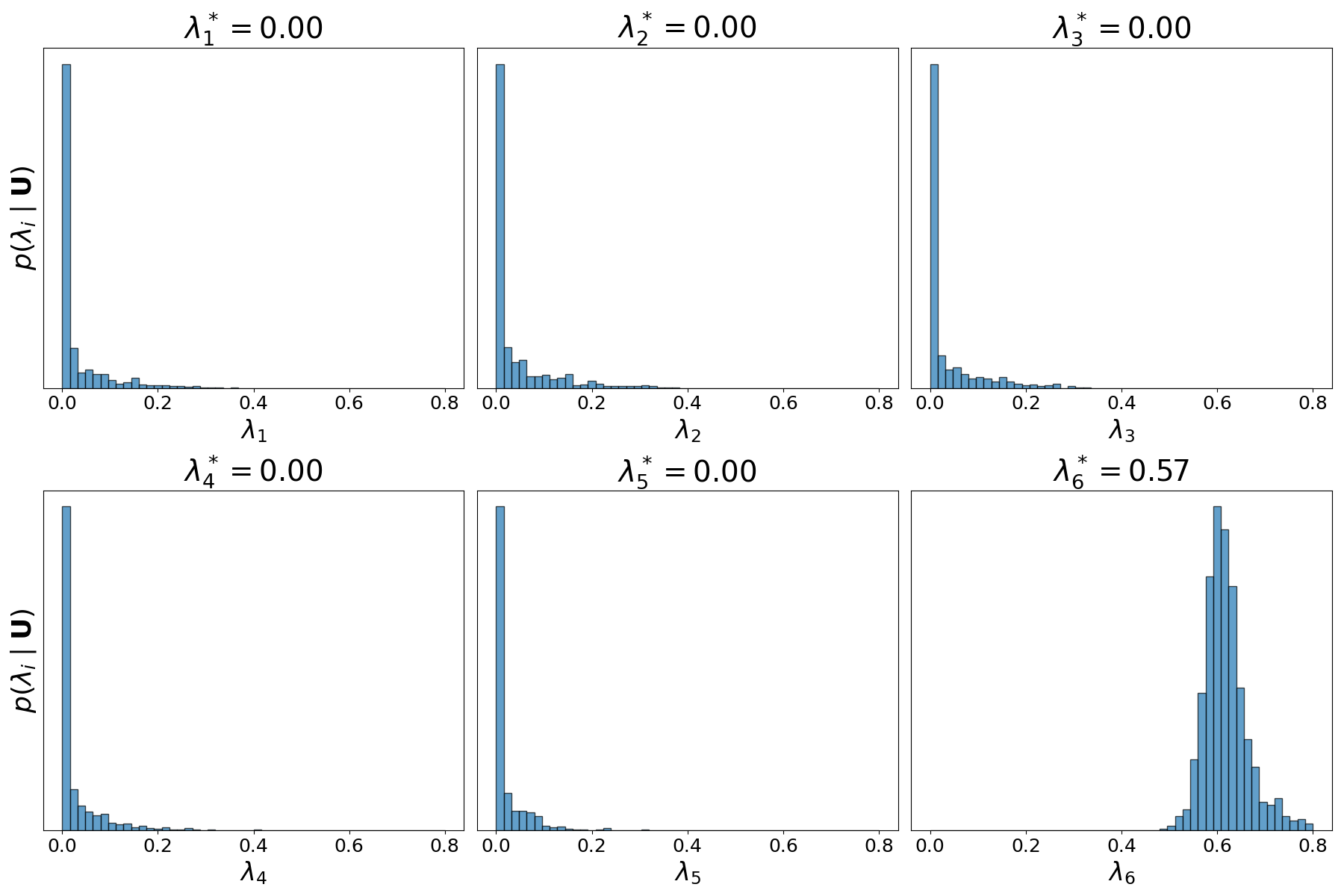}
    \caption{Railway bridge - Exemplary DE-MCMC result showing the posterior marginal distributions from $p(\boldsymbol{\lambda} \mid \mathbf{U})$, each associated with the damage magnitude within a damageable subdomain.}

    \label{fig:bridge-ex}
\end{figure}
 
For visualization purposes, \fig\ref{fig:bridge-ex} presents an exemplary posterior distribution recovered by DE-MCMC for damage located within $\Omega_6$. The parameters support ranges from $0$ to $0.8$, where $0$ corresponds to an undamaged state and values from $0.3$ to $0.8$ indicate stiffness reductions $\delta\in[30\%,80\%]$. Accordingly, the interval $[0, 0.3]$ represents the no-damage regime.

The results obtained from $100$ independent DE-MCMC simulations using the testing dataset $\mathcal{D}^\text{test}_{\boldsymbol{\lambda}}$ are summarized in \tab\ref{tab:meanbridge}. For each one of them, damage identification is performed using two point estimators from the posterior distribution $p(\boldsymbol{\lambda}\mid\mathbf{U})$, namely the posterior mean \(\widehat{\boldsymbol{\lambda}}_{\text{mean}} = \mathbb{E}[p (\boldsymbol{\lambda} \mid\mathbf{U})],\)
and the maximum a posteriori (MAP) estimate
\(\widehat{\boldsymbol{\lambda}}_{\text{MAP}} = \arg\max_{\boldsymbol{\lambda}} \, p(\boldsymbol{\lambda}\mid\mathbf{U}).\) The ground-truth parameter vector used to generate the observations is denoted by $\boldsymbol{\lambda}^*$.

The correct damage class is identified with an accuracy of $93\%$ when using posterior mean and $97\%$ when using MAP. Moreover, the absolute error in the estimated damage magnitude $\delta$ is approximately $6\%$ for the posterior mean and $7\%$ for the posterior mode, over an admissible support of $50\%$. The posterior average standard deviations for each damage class are reported in \tab\ref{tab:stadevbridge}. Overall, both damage location and magnitude are identified with high accuracy and relatively low uncertainty, with no cases exhibiting significant discrepancies between the true damage parameters and the posterior estimates.

\begin{table}[!t]
\centering
\caption{Railway bridge - Damage localization and quantification results under different operational and damage conditions, reported in terms of classification accuracy and absolute error for the posterior mean and posterior mode.}
\begin{tabular}{lcc}
\toprule
\textbf{Metric} & \textbf{With $\widehat{\boldsymbol{\lambda}}_{\text{mean}}$} & \textbf{With $\widehat{\boldsymbol{\lambda}}_{\text{MAP}}$} \\
\hline
Classification accuracy   & 93\% & 97\% \\
Absolute error & 0.062 & 0.072  \\
\bottomrule
\end{tabular}
\label{tab:meanbridge}
\end{table}

\begin{table}[!t]
\centering
\caption{Railway bridge - Posterior standard deviation of the estimated damage magnitude for each damage location.}
\begin{tabular}{l|cccccc}
\toprule
\textbf{True damage region} & $\Omega_1$ & $\Omega_2$ & $\Omega_3$ & $\Omega_4$ & $\Omega_5$ & $\Omega_6$ \\
\hline
\textbf{Standard deviation}  & 0.0787 & 0.0670 & 0.0907 & 0.1196 & 0.1422 & 0.0899 \\
\bottomrule
\end{tabular}
\label{tab:stadevbridge}
\end{table}

\subsection{Groundwater flow case study}
This second case study concerns a groundwater flow problem, adapted from the work of~\cite{lykkegaard2021accelerating}. It considers a steady-state flow in a confined, inhomogeneous aquifer with domain \mbox{$\Omega = (0,1) \times (0,1)$}. Under the assumption of a stationary and incompressible fluid flow, the governing equation that models the pressure distribution $\hh(\boldsymbol{\xx})$ in a porous medium is the following scalar elliptic partial differential equation:
\begin{equation}
\begin{cases}
- \nabla \cdot \left(T(\boldsymbol{\xx}) \nabla \hh(\boldsymbol{\xx}) \right) = g(\boldsymbol{\xx}) &\forall\boldsymbol{\xx}\text{ in }\Omega,\\
\hh(\boldsymbol{\xx}) = \hh_D(\boldsymbol{\xx}) &\forall\boldsymbol{\xx}\text{ on } \Gamma_D, \\
\left(T(\boldsymbol{\xx}) \nabla \hh(\boldsymbol{\xx}) \right) \cdot \mathbf{n} = \qq_N(\boldsymbol{\xx}) &\forall\boldsymbol{\xx}\text{ on } \Gamma_N. 
\end{cases}
\label{eq:flux}
\end{equation}
Here, $T(\boldsymbol{\xx})$ denotes the heterogeneous, depth-integrated transmissivity, while in the absence of internal sources or sinks, $g(\boldsymbol{\xx}) = 0$. 
The function $\hh_D(\boldsymbol{\xx})$ prescribes the hydraulic head on the Dirichlet boundary $\Gamma_D = \{0\}\times(0,1)\,\cup\,\{1\}\times(0,1)$, while $\qq_N(\boldsymbol{\xx})$ denotes the Darcy flux across the Neumann boundary $\Gamma_N = \partial\Omega \setminus \Gamma_D$, with $\mathbf{n}$ being the outward unit normal vector on $\Gamma_N$. 

To approximate the hydraulic head $\hh(\boldsymbol{\xx})$, we introduce a finite-dimensional subspace defined over a computational mesh with $N_N$ nodes, spanned by piecewise linear Lagrange basis functions. Substituting the discrete approximation of the hydraulic head into the weak formulation of problem \eqref{eq:flux} yields a system of sparse linear equations, whose solution vector $\boldsymbol{\mathfrak{h}}\in \mathbb{R}^{N_N}$ collects the nodal values of the hydraulic head.

With the aim of identifying a spatially varying hydraulic conductivity field, the aquifer transmissivity is modeled as a log-Gaussian random field. This modeling choice is well established in the literature~\cite{freeze1975stochastic, kitterrod1997klkriging}, particularly for problems in which the medium does not exhibit strongly correlated extreme values, geological faults, or sharp discontinuities. Accordingly, the spatial correlation structure of the transmissivity field is characterized by a covariance kernel $C(\boldsymbol{\xx}, \boldsymbol{\xx'})$. In our numerical experiments, we adopt a squared exponential (Gaussian) kernel:
\begin{equation}
C(\boldsymbol{\xx}, \boldsymbol{\xx}') = \exp\left( -\frac{1}{2} \sum_{j=1}^{N_D} \left( \frac{\xx_j - \xx'_j}{\ell} \right)^2 \right),
\label{eq:covariance}
\end{equation}
where $N_D$ denotes the spatial dimensionality of the problem and $\ell \in \mathbb{R}^+$ is the correlation length controlling the spatial smoothness of the field.

\begin{figure}[!b]
    \centering
    \includegraphics[width=0.9\linewidth]{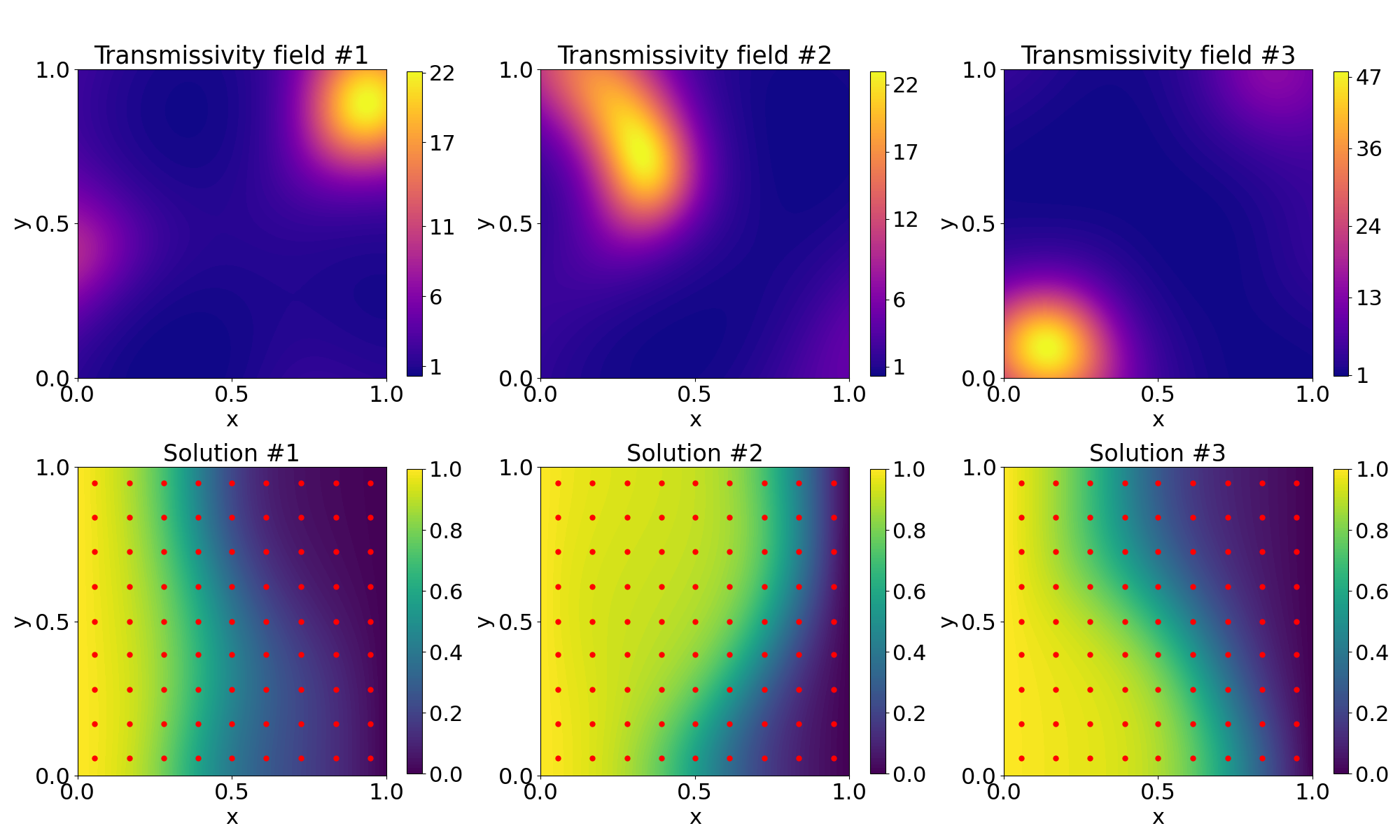}
    \caption{Groundwater flow - Exemplary realizations of the transmissivity field and the corresponding hydraulic head solutions. Sensor locations used for inference are highlighted in red.}
    \label{fig:plot_4}
\end{figure}

To parametrize the random field, we construct the covariance matrix $\mathbf{C} \in \mathbb{R}^{N_N \times N_N}$, with entries $C_{ij} = C(\boldsymbol{\xx}_i, \boldsymbol{\xx}_j)$, $i,j=1,\ldots,N_N$, where $\boldsymbol{\xx}_i$ and $\boldsymbol{\xx}_j$ denote the spatial coordinates of the $i$-th and $j$-th nodes of the finite element mesh, respectively. The random field is then parametrized using a Karhunen-Loève (KL) expansion. Let $\langle\pi_i, \boldsymbol{\psi}_i\rangle$, $i=1,\ldots,N_N$, denote the eigenpairs of the covariance matrix $\mathbf{C}$, sorted in descending order of the eigenvalues. Truncating the expansion after the first $N_M$ dominant modes, the KL representation of the log-transmissivity field reads:
\begin{equation}
\log \mathbf{t} = \boldsymbol{\mu}_\mathbf{t} + \sigma_\mathbf{t} \, \mathbf{\Psi} \, \boldsymbol{\Pi}^{1/2} \, \boldsymbol{\lambda},
\end{equation}
where $\mathbf{t}\in\mathbb{R}^{N_N}$ is the discretized transmissivity field, $\boldsymbol{\mu}_\mathbf{t} \in \mathbb{R}^{N_N}$ denotes the mean vector of the log-transmissivity field, $\sigma_ \mathbf{t} \in \mathbb{R}^+$ is its marginal standard deviation, $\boldsymbol{\Psi} = [\boldsymbol{\psi}_1, \ldots, \boldsymbol{\psi}_{N_M}]\in\mathbb{R}^{N_N\times N_M}$ collects as columns the retained eigenvectors, $\boldsymbol{\Pi} = \text{diag}\left[\pi_1,\ldots,\pi_{N_M}\right]$ is a diagonal matrix collecting the corresponding eigenvalues, and $\boldsymbol{\lambda} \sim \mathcal{N}(\mathbf{0}_{N_M},\mathbf{I}_{N_M})$ is a vector of $N_\text{par}=N_M$ independent standard Gaussian random variables. This latter provides a low-dimensional representation of the spatially varying transmissivity field and constitutes the set of target parameters to be inferred.

The computational domain $\Omega$ is discretized using linear finite elements with a dimensionless element size of $1/60$. A Dirichlet unit hydraulic head is prescribed on the left boundary, while a homogeneous Dirichlet condition is imposed on the right boundary. Zero-flux Neumann boundary conditions are imposed on the top and bottom boundaries.

The log-transmissivity field is modeled with unit mean $\boldsymbol{\mu}_\mathbf{t}$ and unit marginal standard deviation $\sigma_\mathbf{t}$. The spatial correlation length is set to $\ell = 0.25$. The resulting covariance matrix is decomposed via KL expansion and truncated after the first $N_M=14$ dominant modes, which collectively retain approximately $97\%$ of the total variance. Accordingly, the log-transmissivity field is parametrized by a low-dimensional vector of $N_\text{par}=N_M=14$ independent Gaussian coefficients $\boldsymbol{\lambda}$. Figure~\ref{fig:plot_4} illustrates exemplary realizations of the transmissivity field together with the corresponding hydraulic head solutions $\boldsymbol{\mathfrak{h}}$.

Inference over the parameter vector $\boldsymbol{\lambda}$ is performed using hydraulic head measurements collected on a regular $9 \times 9$ grid of equispaced sensor locations, highlighted in \fig\ref{fig:plot_4}. These measurements are extracted from the full hydraulic head solution $\boldsymbol{\mathfrak{h}}$ through a Boolean selection operator and constitute the raw observation vector. $\mathbf{x}$, with dimension $N_x=81$. With this setup, the training dataset $\mathcal{D}^\text{train}_{\boldsymbol{\lambda}}$ is populated with $32,000$ realizations to train the DL models, while the testing dataset $\mathcal{D}^\text{test}_{\boldsymbol{\lambda}}$ is populated with $8000$ additional instances.

\begin{figure}[!t]
    \centering
    \includegraphics[width=1\linewidth]{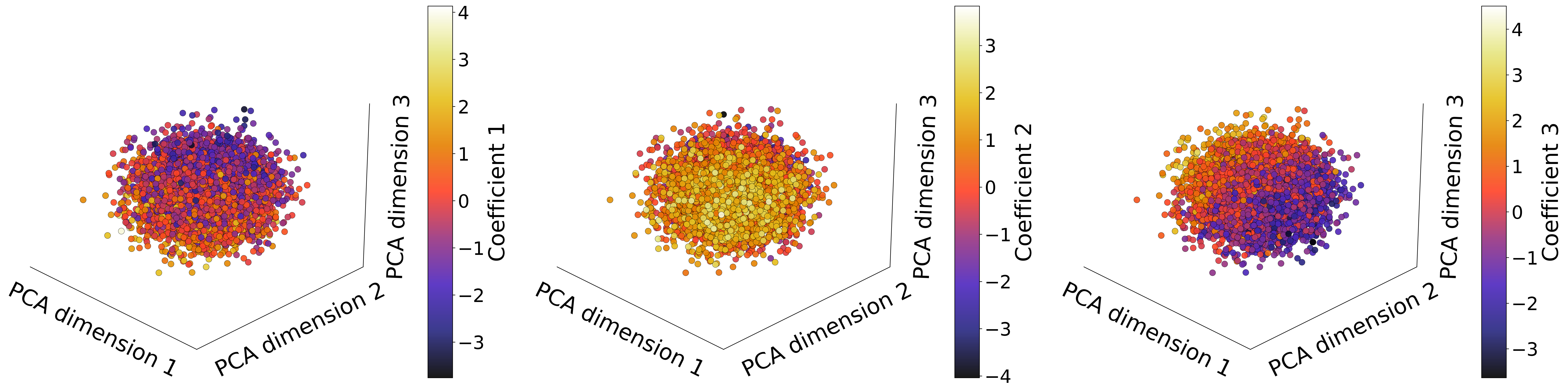}
    \caption{Groundwater flow - Three-dimensional principal component projections of the latent-space mean vectors for the testing data. From left to right, the color coding corresponds to the true values of the first three coefficients of the Karhunen-Loève representation of the log-transmissivity field.}
    \label{fig:pca-flow}
\end{figure}

A three-dimensional PCA projection of the $N_h=20$-dimensional latent mean vectors extracted by the i-VAE for each instance in $\mathcal{D}^\text{test}_{\boldsymbol{\lambda}}$ is shown in \fig\ref{fig:pca-flow}. The color coding corresponds to the values of the first three coefficients of the KL representation of the log-transmissivity field. Compared to the previous case study, the resulting low-dimensional representation appears fuzzier, with only weakly discernible structure and no clear manifold topology. This behavior can be attributed to the increased complexity of the mapping between the parameter vector $\boldsymbol{\lambda}$ and the observations $\mathbf{x}$, as well as to the higher dimensionality of $\boldsymbol{\lambda}$ in this problem.

\begin{figure}[!t]
    \centering
    \includegraphics[width=0.7\linewidth]{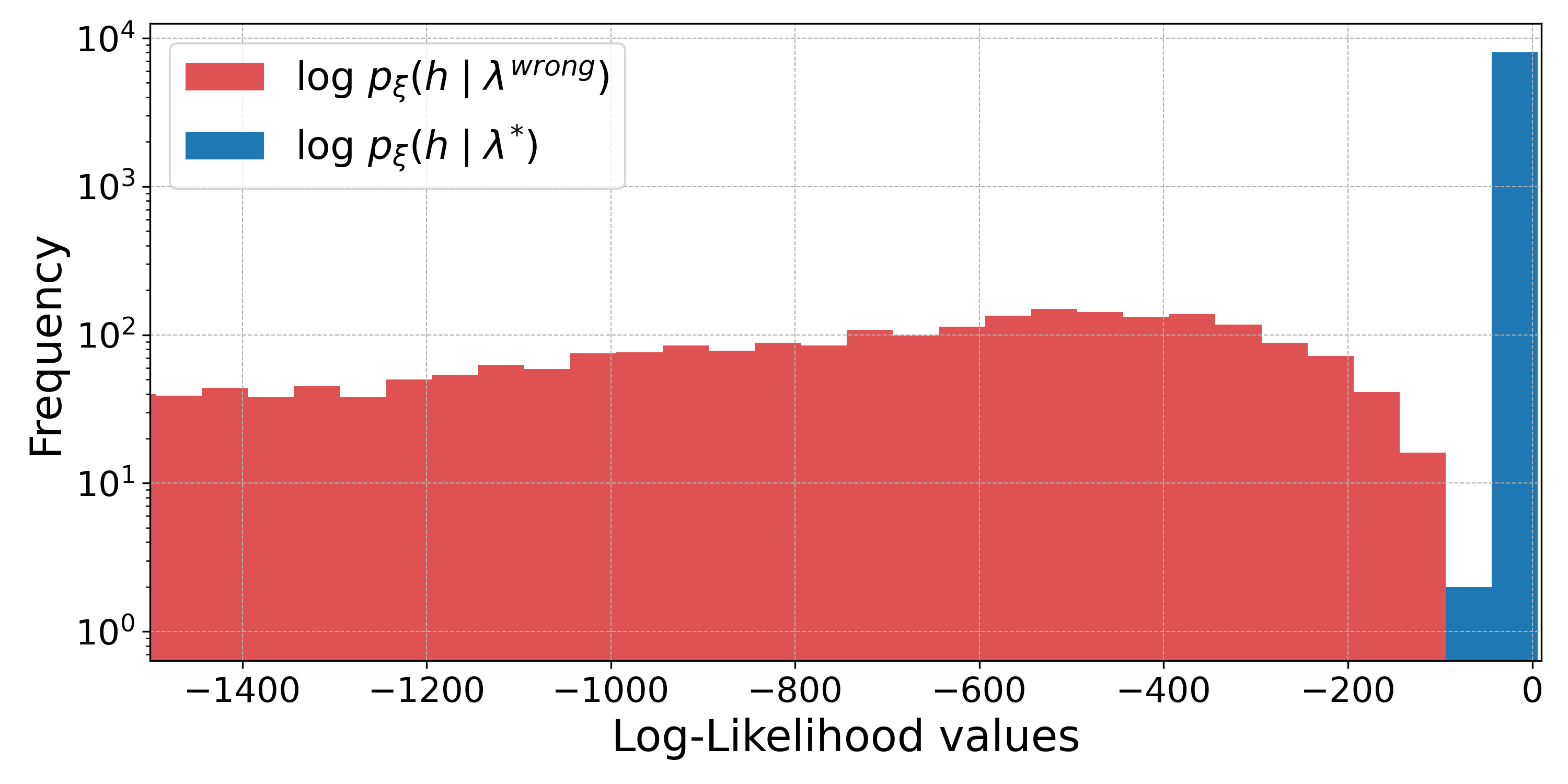}
    \caption{Groundwater flow - Frequency distribution of the log-likelihood values for latent representations paired with correct labels (blue) and incorrect labels (red).}
    \label{fig:loglikflow}
\end{figure}

The ability of the CNF to assign informative log-likelihoods values is summarized in \fig\ref{fig:loglikflow}. As in the previous case study, the CNF is evaluated on each test sample twice: once paired with its true coefficient vector $\boldsymbol{\lambda}$, shown in blue, and once paired with a randomly selected parameter vector drawn from the testing dataset $\mathcal{D}^\text{test}_{\boldsymbol{\lambda}}$, shown in red. The resulting distribution clearly indicates that the CNF consistently assigns higher log-likelihood values to correctly labeled samples, while penalizing incorrect parameterizations. This behavior confirms the ability of the CNF to reliably discriminate between compatible and incompatible $(\mathbf{x}, \boldsymbol{\lambda})$ pairs, even in the presence of a high-dimensional nonlinear inverse mapping.

\begin{figure}[t!]
    \centering    \includegraphics[width=0.95\linewidth]{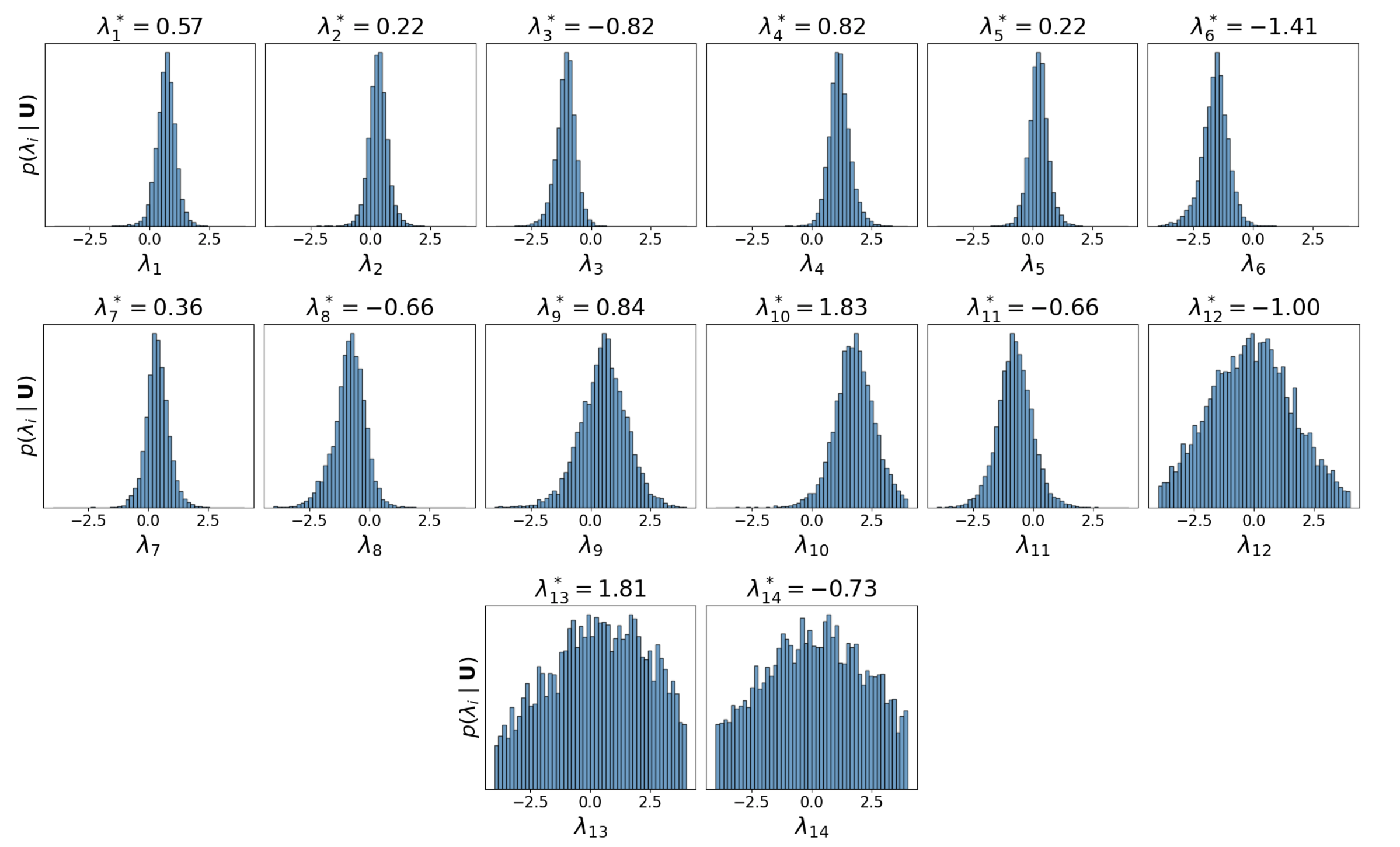}
    \caption{Groundwater flow - Exemplary MCMC result showing the posterior distributions of the $14$ coefficients of the Karhunen-Loève representation of the log-transmissivity field.}
    \label{fig:output-flow}
\end{figure}

Starting from a standard multivariate normal prior $p(\boldsymbol{\lambda})$, the posterior distribution $p(\boldsymbol{\lambda}\mid\mathbf{x})$ is approximated using the DE-MCMC algorithm. Observed hydraulic head measurements $\mathbf{x}$ are first mapped into a low-dimensional latent space by the i-VAE encoder, producing a probabilistic latent representation $\mathbf{h}\sim\mathcal{N}(\boldsymbol{\mu}_\theta,\, \boldsymbol{\sigma}^2_\theta)$. Joint samples $\langle\mathbf{h}' ,\boldsymbol{\lambda}'\rangle$ are then drawn from the latent and parameter spaces, and the CNF is used to evaluate the conditional likelihood $p(\mathbf{h}' \mid \boldsymbol{\lambda}')$, which efficiently informs the DE-MCMC acceptance ratio.

\begin{figure}[!t]
    \centering
\includegraphics[width=0.8\linewidth]{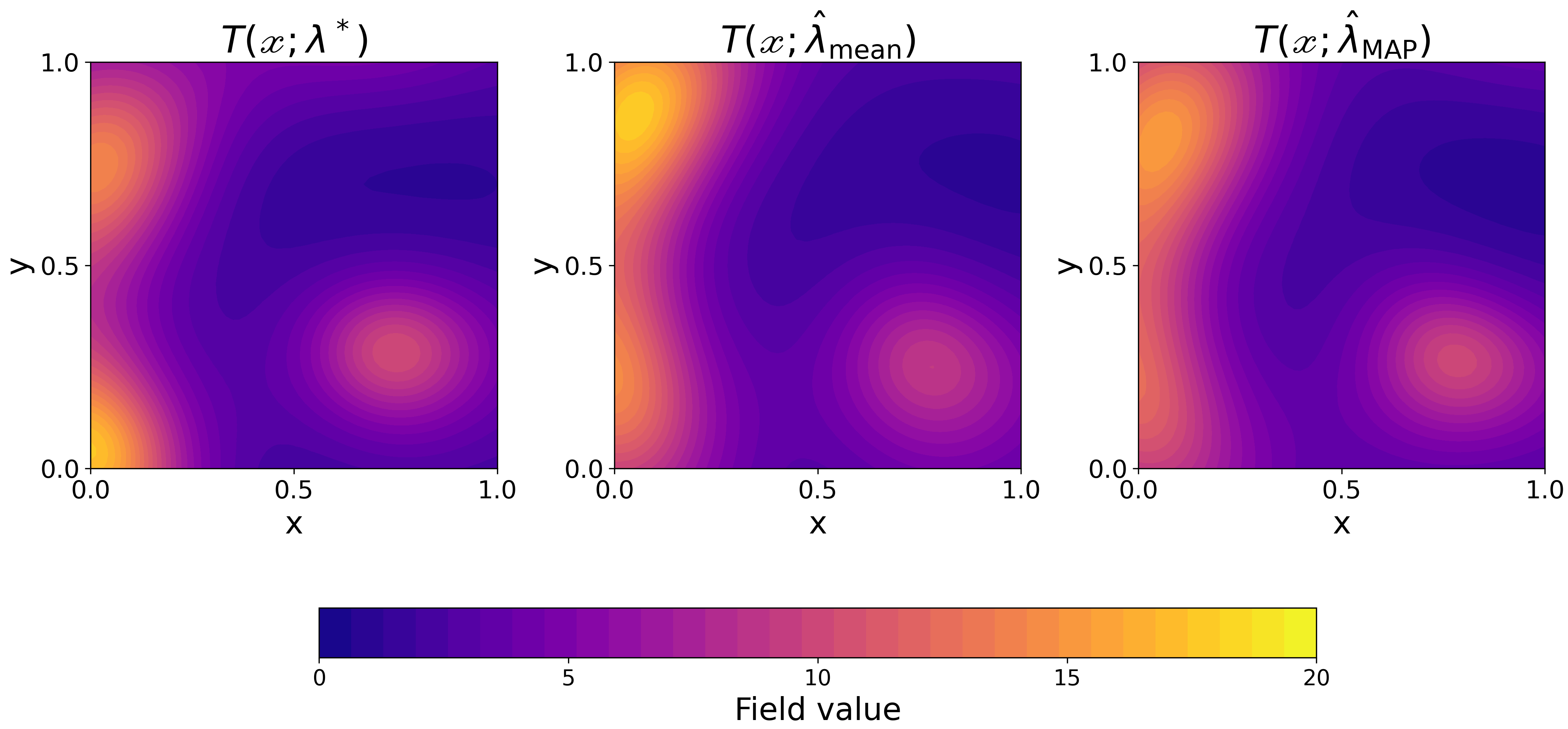}
    \caption{Groundwater flow - Exemplary comparison between (left) ground truth, (middle) estimated transmissivity fields using the posterior mean of the Karhunen-Loève coefficients, and (right)  estimated transmissivity fields using the posterior mode of the Karhunen-Loève coefficients. }
    \label{fig:contour_fields}
\end{figure}

Two chains are run sequentially, each with $N_\text{s} = 30,000$ iterations, ensuring a Gelman-Rubin statistic below 1.01 in all cases. The runtime for each chain is approximately 3 minutes. An exemplary MCMC output is shown in \fig\ref{fig:output-flow}. The corresponding transmissivity field, reconstructed using either the posterior mean or posterior mode of the KL coefficients, is compared with the ground truth in \fig\ref{fig:contour_fields}.

\begin{figure}[!t]
    \centering
    \includegraphics[width=0.7\linewidth]{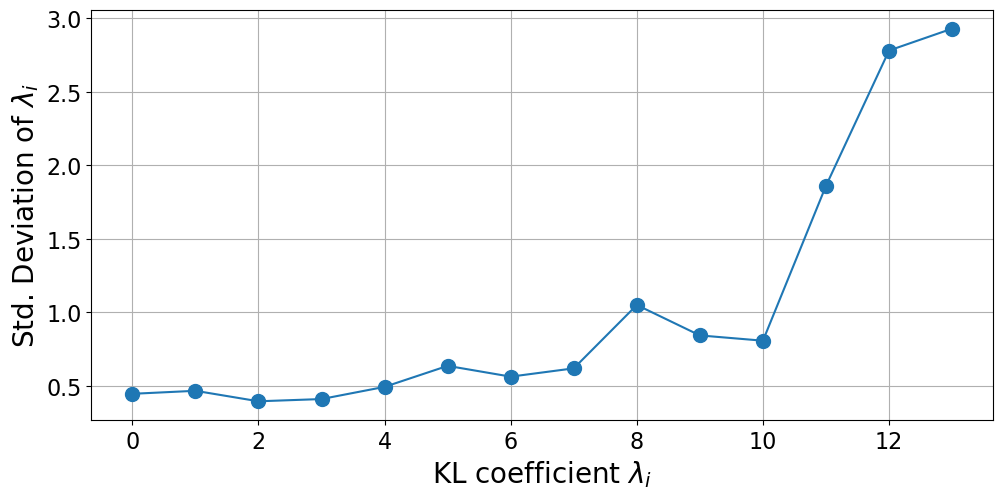}
    \caption{Groundwater flow - Posterior standard deviation of the Karhunen–Loève coefficients, averaged over the testing dataset $\mathcal{D}^\text{test}_{\boldsymbol{\lambda}}$. }
    \label{fig:std_dev_trend}
\end{figure}

The increasing uncertainty observed in the posterior distributions from the first to the last KL coefficients is documented in \fig\ref{fig:std_dev_trend}. This trend is expected, since higher-order KL modes contribute less to the overall structure of the transmissivity field and are therefore less constrained by the observations. However, this increased uncertainty has a negligible impact on the accuracy of the reconstructed transmissivity fields, as the influence of these modes on the solution is limited.

\begin{figure}[!t]
    \centering
    \includegraphics[width=0.7\linewidth]{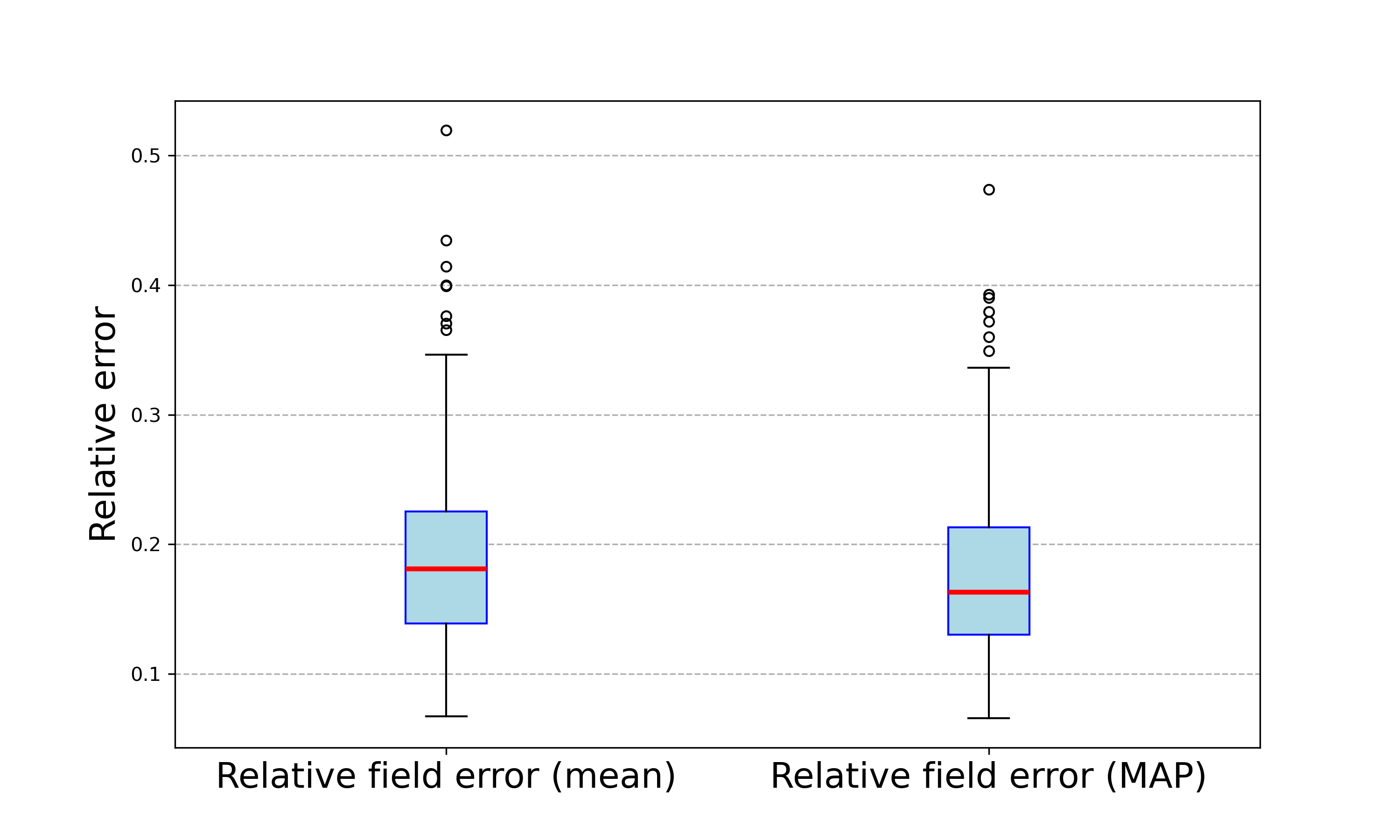}
    \caption{Groundwater flow - Box plots of the relative errors in transmissivity identification for $100$ random samples from $\mathcal{D}^\text{test}_{\boldsymbol{\lambda}}$, obtained from field reconstructions using (left) the posterior mean and (right) the MAP of the Karhunen-Loève coefficients.}
    \label{fig:errors}
\end{figure}

To quantify the reconstruction performance, 100 DE-MCMC simulations are performed on randomly selected test samples from $\mathcal{D}^\text{test}_{\boldsymbol{\lambda}}$. The relative error is computed as
\begin{equation}
        \text{relative error}=\frac{\lVert\mathbf{t}_{\text{true}} - \mathbf{t}_{\text{pred}}\rVert_{2}}{\lVert\mathbf{t}_{\text{true}}\rVert_{2}}.
\end{equation}
The reconstructed transmissivity fields using the posterior mean and posterior mode achieve relative errors of approximately $0.186$ and $0.168$, respectively; see also \fig\ref{fig:errors} for the corresponding box plots. These results demonstrate that the i-VAE provides a compact and informative latent representation, enabling the CNF to produce accurate log-likelihoods, which in turn supports reliable posterior inference via DE-MCMC.

\section{Conclusions}
\label{sec:conclusion}

This work introduces a simulation-based inference framework for inverse problems, integrating representation learning, conditional density estimation, and Bayesian sampling. The proposed methodology embeds a supervised variational autoencoder (VAE) and a conditional normalizing flow (CNF) within a Markov chain Monte Carlo (MCMC) scheme, enabling efficient posterior inference in high-dimensional and nonlinear settings.

A key contribution is the design of a modified VAE that learns a probabilistic latent representation explicitly informed by the parameters of interest. By augmenting the standard VAE architecture with a prediction branch operating on latent samples, the resulting latent space is both regularized and parameter-informative, providing a compact and uncertainty-aware summary of the observations. A CNF trained in this latent space approximates the conditional likelihood of the latent features given the parameters, replacing repeated forward model evaluations during MCMC sampling while preserving a fully Bayesian treatment of uncertainty.

The framework has been demonstrated on two distinct inverse problems: damage localization and quantification in a railway bridge, and identification of a spatially varying hydraulic transmissivity field in a groundwater flow problem. In both cases, the approach achieved accurate parameter inference and meaningful uncertainty quantification, despite the high dimensionality and ill-posedness of the problems.

The proposed framework is broadly applicable to inverse problems involving expensive forward models and limited observational data. Future work will focus on extensions to time-dependent settings, adaptive learning strategies, and tighter integration within digital twin architectures for real-time data assimilation and decision support~\cite{MT_AIF}.

\vspace{6pt} 
\noindent{\bf Competing Interests:} The authors declare that they have no known competing financial interests or personal relationships that could have appeared to influence the work reported in this paper.

\vspace{6pt} 
\noindent{\bf Acknowledgments:} The authors thank Dr.~Filippo Zacchei (MOX-Department of Mathematics, Politecnico di Milano) for his insightful comments and remarks.

\vspace{6pt} 
\noindent{\bf Funding:} 
GB acknowledges the support of European Union - NextGenerationEU within the Italian PNRR program (M4C2, Investment 3.3) for the PhD Scholarship ‘‘Physics-informed Artificial Intelligence for surrogate modeling of complex energy systems’’.

GB and AM are members of the Gruppo Nazionale Calcolo Scientifico-Istituto Nazionale di Alta Matematica (GNCS-INdAM). AM acknowledges the project ‘‘Dipartimento di Eccellenza’’ 2023–2027, funded by MUR, as well as the financial support from the FIS starting grant DREAM (Grant Agreement FIS00003154), funded by the Italian Science Fund (FIS) - Ministero dell’Università e della Ricerca.  

MT and AM acknowledge the financial support from the ERC advanced grant IMMENSE (Grant Agreement 101140720), funded by the European Union. Views and opinions expressed are however those of the authors only and do not necessarily reflect those of the European Union or the European Research Council Executive Agency. Neither the European Union nor the granting authority can be held responsible for them.

\medskip

\bibliographystyle{elsarticle-num}
\biboptions{sort&compress}

\appendix
\section{Implementation details of the deep learning models} 
\label{sec:implementation}

In this Appendix, we describe the implementation details of the deep learning models employed in \sez\ref{sec:3} to validate the proposed inference framework. The considered two case studies deal with the structural health monitoring of a railway bridge and the identification of a spatially varying transmissivity field in a groundwater flow problem. Although these problems involve different physical systems and observation models, they are both addressed using a common learning framework that combines latent-variable representation with conditional density estimation.

The architecture, hyperparameters, and training settings of the informed-VAE --- composed of the encoder, decoder, and latent-space predictor --- are summarized in \tab\ref{tab:NN_VAE_arch_1} for the railway bridge case study and in \tab\ref{tab:NN_VAE_arch_2} for the groundwater flow case study. The encoder consists of a stack of convolutional layers interleaved with max-pooling and dropout operations, followed by fully connected layers that map the input data onto a latent space of dimension $N_h$. The decoder mirrors the encoding process through fully connected layers and transposed convolutional layers combined with upsampling operations, reconstructing the input from the latent representation. The latent-space predictor consists of three fully-connected layers.
 
The RealNVP-based conditional normalizing flow architectures, together with the corresponding hyperparameters and training settings, are reported in \tab\ref{tab:CNF_1} and \tab\ref{tab:CNF_2} for the railway bridge case and groundwater flow case studies, respectively. The scaling and translation subnetworks, as well as the conditioning network, are implemented as fully connected neural networks. 

Training is performed by minimizing either $\mathcal{L}_{\text{i-VAE}}$ or $\mathcal{L}_{\text{CNF}}$ for the informed-VAE and the conditional normalizing flow, respectively, using the Adam optimizer~\cite{art:Adam}. In both cases, the learning rate is reduced over the last $4/5$ of the total training steps according to a cosine decay schedule. The optimization uses an $80:20$ split of the dataset for training and validation, employing an early stopping strategy to prevent overfitting.

\begin{table}[H]
\caption{Railway bridge - informed-VAE: (a) employed architecture and (b) selected hyperparameters and training settings.\label{tab:NN_VAE_arch_1}}
\centering

\subfloat[]{
\scriptsize
\begin{tabular}{lllll}
\toprule
\mbox{Layer} & \mbox{Type} & \mbox{Output shape} & \mbox{Activ.} & \mbox{Input layer}\\
\toprule
\multicolumn{5}{c}{\textbf{Encoder}}\\
\toprule
0  & Input & $(B_\text{VAE}, L, N_s)$ & None & None\\
1  & Conv1D & $(B_\text{VAE}, L, 32)$ & Tanh & 0\\
2  & MaxPool1D & $(B_\text{VAE}, L/2, 32)$ & None & 1\\
3  & Dropout & $(B_\text{VAE}, L/2, 32)$ & None & 2\\
4  & Conv1D & $(B_\text{VAE}, L/2, 64)$ & Tanh & 3\\
5  & MaxPool1D & $(B_\text{VAE}, L/4, 64)$ & None & 4\\
6  & Dropout & $(B_\text{VAE}, L/4, 64)$ & None & 5\\
7  & Conv1D & $(B_\text{VAE}, L/4, 32)$ & Tanh & 6\\
8  & MaxPool1D & $(B_\text{VAE}, L/8, 32)$ & None & 7\\
9  & Dropout & $(B_\text{VAE}, L/8, 32)$ & None & 8\\
10 & Flatten & $(B_\text{VAE}, 32\cdot L/8)$ & None & 9\\
11 & Dense & $(B_\text{VAE}, 64)$ & Tanh & 10\\
12 & Dense & $(B_\text{VAE}, 16)$ & Tanh & 11\\
13 & Dense ($\boldsymbol{\mu}_\theta$) & $(B_\text{VAE}, N_h)$ & None & 12\\
14 & Dense ($\log\boldsymbol{\sigma}_\theta^2$) & $(B_\text{VAE}, N_h)$ & None & 12\\
15 & Sampling & $(B_\text{VAE}, N_h)$ & None & 13,14\\
\toprule
\multicolumn{5}{c}{\textbf{Decoder}}\\
\toprule
16 & Dense & $(B_\text{VAE}, 16)$ & Tanh & 15\\
17 & Dense & $(B_\text{VAE}, 64)$ & Tanh & 16\\
18 & Dense & $(B_\text{VAE}, 32\cdot L/8)$ & Tanh & 17\\
19 & Reshape & $(B_\text{VAE}, L/8, 32)$ & None & 18\\
20 & Conv1D$^\top$ & $(B_\text{VAE}, L/8, 32)$ & Tanh & 19\\
21 & UpSampling1D & $(B_\text{VAE}, L/4, 32)$ & None & 20\\
22 & Dropout & $(B_\text{VAE}, L/4, 32)$ & None & 21\\
23 & Conv1D$^\top$ & $(B_\text{VAE}, L/4, 64)$ & Tanh & 22\\
24 & UpSampling1D & $(B_\text{VAE}, L/2, 64)$ & None & 23\\
25 & Dropout & $(B_\text{VAE}, L/2, 64)$ & None & 24\\
26 & Conv1D$^\top$ & $(B_\text{VAE}, L/2, 32)$ & Tanh & 25\\
27 & UpSampling1D & $(B_\text{VAE}, L, 32)$ & None & 26\\
28 & Dropout & $(B_\text{VAE}, L, 32)$ & None & 27\\
29 & Conv1D & $(B_\text{VAE}, L, N_s)$ & Linear & 28\\
\toprule
\multicolumn{5}{c}{\textbf{Latent-space Predictor}}\\
\toprule
30 & Dense & $(B_\text{VAE}, 64)$ & Tanh & 15\\
31 & Dense & $(B_\text{VAE}, 32)$ & Tanh & 30\\
32 & Dense & $(B_\text{VAE}, N_{\text{par}})$ & Linear & 31\\
\bottomrule
\end{tabular}
}

\subfloat[]{
\scriptsize
\begin{tabular}{ll}
\toprule
\mbox{Latent dimension:} & $N_h = 4$\\
\mbox{Conv kernel sizes:} & $25,\;13,\;7$\\
\mbox{Dropout rate:} & $0.05$\\
\mbox{Weight initializer:} & \mbox{Xavier}\\
\mbox{$L^2$ regularization rate:} & $10^{-3}$\\
\mbox{Optimizer:} & Adam\\
\mbox{Batch size:} & $B_\text{VAE}=32$\\
\mbox{Initial learning rate:} & $10^{-3}$\\
\mbox{Max epochs:} & $250$\\
\mbox{Learning schedule:} & $0.8$ cosine decay\\
\mbox{Weight decay:} & \mbox{$0.05$}\\
\mbox{Early stopping patience:} & $20$ epochs\\
\mbox{Train--validation split:} & $80{:}20$\\
\mbox{KL weight:} & $\beta_{\text{KL}}=1$\\
\mbox{Prediction weight:} & $\beta_{\text{pred}}=15000$\\
\bottomrule
\end{tabular}
}
\end{table}

\begin{table}[H]
\caption{Groundwater flow - informed-VAE: (a) employed architecture and (b) selected hyperparameters and training settings.\label{tab:NN_VAE_arch_2}}
\centering

\subfloat[]{
\scriptsize
\begin{tabular}{lllll}
\toprule
\mbox{Layer} & \mbox{Type} & \mbox{Output shape} & \mbox{Activ.} & \mbox{Input layer}\\
\toprule
\multicolumn{5}{c}{\textbf{Encoder}}\\
\toprule
0  & Input & $(B_\text{VAE}, N_x)$ & None & None\\
1  & Reshape & $(B_\text{VAE}, \sqrt{N_x}, \sqrt{N_x}, 1)$ & None & 0\\
2  & Conv2D & $(B_\text{VAE}, \sqrt{N_x}, \sqrt{N_x}, 16)$ & Tanh & 1\\
3  & Dropout & $(B_\text{VAE}, \sqrt{N_x}, \sqrt{N_x}, 16)$ & None & 2\\
4  & Conv2D & $(B_\text{VAE}, \sqrt{N_x}, \sqrt{N_x}, 32)$ & Tanh & 3\\
5  & Dropout & $(B_\text{VAE}, \sqrt{N_x}, \sqrt{N_x}, 32)$ & None & 4\\
6  & Conv2D & $(B_\text{VAE}, \sqrt{N_x}, \sqrt{N_x}, 16)$ & Tanh & 5\\
7  & Dropout & $(B_\text{VAE}, \sqrt{N_x}, \sqrt{N_x}, 16)$ & None & 6\\
8  & Flatten & $(B_\text{VAE}, 16N_x)$ & None & 7\\
9  & Dense & $(B_\text{VAE}, 32)$ & Tanh & 8\\
10 & Dropout & $(B_\text{VAE}, 32)$ & None & 9\\
11 & Dense ($\boldsymbol{\mu}_\theta$) & $(B_\text{VAE}, N_h)$ & None & 10\\
12 & Dense ($\log\boldsymbol{\sigma}_\theta^2$) & $(B_\text{VAE}, N_h)$ & None & 10\\
13 & Sampling & $(B_\text{VAE}, N_h)$ & None & 11,12\\
\toprule
\multicolumn{5}{c}{\textbf{Decoder}}\\
\toprule
14 & Input & $(B_\text{VAE}, N_h)$ & None & 13\\
15 & Dense & $(B_\text{VAE}, 32)$ & Tanh & 14\\
16 & Dense & $(B_\text{VAE}, 16N_x)$ & Tanh & 15\\
17 & Reshape & $(B_\text{VAE}, \sqrt{N_x}, \sqrt{N_x}, 16)$ & None & 16\\
18 & Conv2D$^\top$ & $(B_\text{VAE}, \sqrt{N_x}, \sqrt{N_x}, 32)$ & Tanh & 17\\
19 & Dropout & $(B_\text{VAE}, \sqrt{N_x}, \sqrt{N_x}, 32)$ & None & 18\\
20 & Conv2D$^\top$ & $(B_\text{VAE}, \sqrt{N_x}, \sqrt{N_x}, 16)$ & Tanh & 19\\
21 & Dropout & $(B_\text{VAE}, \sqrt{N_x}, \sqrt{N_x}, 16)$ & None & 20\\
22 & Conv2D$^\top$ & $(B_\text{VAE}, \sqrt{N_x}, \sqrt{N_x}, 1)$ & Tanh & 21\\
23 & Dropout & $(B_\text{VAE}, \sqrt{N_x}, \sqrt{N_x}, 1)$ & None & 22\\
24 & Flatten & $(B_\text{VAE}, N_x)$ & None & 23\\
25 & Dense & $(B_\text{VAE}, N_x)$ & Linear & 24\\
\toprule
\multicolumn{5}{c}{\textbf{Latent-space Predictor}}\\
\toprule
26 & Input & $(B_\text{VAE}, N_h)$ & None & 13\\
27 & Dense & $(B_\text{VAE}, 64)$ & Tanh & 26\\
28 & Dropout & $(B_\text{VAE}, 64)$ & None & 27\\
29 & Dense & $(B_\text{VAE}, 32)$ & Tanh & 28\\
30 & Dense & $(B_\text{VAE}, N_{\text{par}})$ & Linear & 29\\
\bottomrule
\end{tabular}
}

\subfloat[]{
\scriptsize
\begin{tabular}{ll}
\toprule
\mbox{Latent dimension:} & $N_h = 20$\\
\mbox{Conv kernel sizes:} & $3\times3$\\
\mbox{Dropout rate:} & $0.2$\\
\mbox{Weight initializer:} & \mbox{Xavier}\\
\mbox{$L^2$ regularization rate:} & $10^{-3}$\\
\mbox{Optimizer:} & Adam\\
\mbox{Batch size:} & $B_\text{VAE}=32$\\
\mbox{Initial learning rate:} & $10^{-3}$\\
\mbox{Max epochs:} & $250$\\
\mbox{Learning schedule:} & $0.8$ cosine decay\\
\mbox{Weight decay:} & \mbox{$0.05$}\\
\mbox{Early stopping patience:} & $25$ epochs\\
\mbox{Train--validation split:} & $80{:}20$\\
\mbox{KL weight:} & $\beta_{\text{KL}}=2.5^{-4}$\\
\mbox{Prediction weight:} & $\beta_{\text{pred}}=10^{-4}$\\
\bottomrule
\end{tabular}
}
\end{table}

\begin{table}[H]
\caption{Railway bridge - conditional normalizing flow: (a) employed architecture and (b) selected hyperparameters and training settings.\label{tab:CNF_1}}
\centering

\subfloat[]{
\scriptsize
\begin{tabular}{lllll}
\toprule
\mbox{Layer} & \mbox{Type} & \mbox{Output shape} & \mbox{Activ.} & \mbox{Input layer}\\
\toprule
0 & Input ($\mathbf{h}$) & $(B_\text{NF}, N_h)$ & None & None\\
1 & Input ($\boldsymbol{\lambda}$) & $(B_\text{NF}, N_{\text{par}})$ & None & None\\
\midrule
\multicolumn{5}{c}{\textbf{Parameter conditioning network}}\\
\midrule
2 & Dense & $(B_\text{NF}, 32)$ & ReLU & 1\\
3 & Dense & $(B_\text{NF}, 64)$ & ReLU & 2\\
4 & Dense & $(B_\text{NF}, 128)$ & ReLU & 3\\
\midrule
\multicolumn{5}{c}{\textbf{Translation network}}\\
\midrule
5 & Dense & $(B_\text{NF}, 64)$ & ReLU & 0\\
6 & Dropout & $(B_\text{NF}, 64)$ & None & 5\\
7 & Dense & $(B_\text{NF}, 128)$ & ReLU & 4\\
8 & Concatenate & $(B_\text{NF}, 192)$ & None & 6,7\\
9 & Dense & $(B_\text{NF}, 64)$ & ReLU & 8\\
10 & Dense & $(B_\text{NF}, N_h)$ & Linear & 9\\
\midrule
\multicolumn{5}{c}{\textbf{Scaling network}}\\
\midrule
11 & Dense & $(B_\text{NF}, 64)$ & ReLU & 0\\
12 & Dropout & $(B_\text{NF}, 64)$ & None & 11\\
13 & Dense & $(B_\text{NF}, 128)$ & ReLU & 4\\
14 & Concatenate & $(B_\text{NF}, 192)$ & None & 12,13\\
15 & Dense & $(B_\text{NF}, 64)$ & ReLU & 14\\
16 & Dense & $(B_\text{NF}, N_h)$ & Tanh & 15\\
\bottomrule
\end{tabular}
}

\subfloat[]{
\scriptsize
\begin{tabular}{ll}
\toprule
\mbox{Latent dimension:} & $N_h = 4$\\
\mbox{Parameters dimension:} & $N_{\text{par}}=6$\\
\mbox{Coupling layers:} & $N_f = 16$\\
\mbox{Masking strategy:} & $N_h/2$ frozen components (alternating)\\
\mbox{Dropout rate:} & $0.1$\\
\mbox{Weight initializer:} & \mbox{Xavier}\\
\mbox{$L^2$ regularization rate:} & $10^{-3}$\\
\mbox{Optimizer:} & Adam\\
\mbox{Batch size:} & $B_\text{NF} = 32$\\
\mbox{Initial learning rate:} & $10^{-3}$\\
\mbox{Max epochs:} & $250$\\
\mbox{Learning schedule:} & $0.8$ cosine decay\\
\mbox{Weight decay:} & \mbox{$0.05$}\\
\mbox{Early stopping patience:} & $15$ epochs\\
\mbox{Train--validation split:} & $80{:}20$\\
\bottomrule
\end{tabular}
}
\end{table}

\begin{table}[H]
\caption{Groundwater flow - conditional normalizing flow: (a) employed architecture and (b) selected hyperparameters and training settings.\label{tab:CNF_2}}
\centering

\subfloat[]{
\scriptsize
\begin{tabular}{lllll}
\toprule
\mbox{Layer} & \mbox{Type} & \mbox{Output shape} & \mbox{Activ.} & \mbox{Input layer}\\
\toprule
0 & Input ($\mathbf{x}$) & $(B_\text{NF}, N_h)$ & None & None\\
1 & Input ($\boldsymbol{\lambda}$) & $(B_\text{NF}, N_{\text{par}})$ & None & None\\
\midrule
\multicolumn{5}{c}{\textbf{Parameter conditioning network}}\\
\midrule
2 & Dense & $(B_\text{NF}, 32)$ & ELU & 1\\
3 & Dropout & $(B_\text{NF}, 32)$ & None & 2\\
4 & Dense & $(B_\text{NF}, 64)$ & ELU & 3\\
5 & Dropout & $(B_\text{NF}, 64)$ & None & 4\\
6 & Dense & $(B_\text{NF}, 128)$ & ELU & 5\\
\midrule
\multicolumn{5}{c}{\textbf{Translation network}}\\
\midrule
7 & Dense & $(B_\text{NF}, 64)$ & ELU & 0\\
8 & Dropout & $(B_\text{NF}, 64)$ & None & 7\\
9 & Dense & $(B_\text{NF}, 64)$ & ELU & 6\\
10 & Concatenate & $(B_\text{NF}, 128)$ & None & 8,9\\
11 & Dense & $(B_\text{NF}, 32)$ & ELU & 10\\
12 & Dense & $(B_\text{NF}, N_h)$ & Linear & 11\\
\midrule
\multicolumn{5}{c}{\textbf{Scaling network}}\\
\midrule
13 & Dense & $(B_\text{NF}, 64)$ & ELU & 0\\
14 & Dropout & $(B_\text{NF}, 64)$ & None & 13\\
15 & Dense & $(B_\text{NF}, 64)$ & ELU & 6\\
16 & Concatenate & $(B_\text{NF}, 128)$ & None & 14,15\\
17 & Dense & $(B_\text{NF}, 32)$ & ELU & 16\\
18 & Dense & $(B_\text{NF}, N_h)$ & Tanh & 17\\
\bottomrule
\end{tabular}
}

\subfloat[]{
\scriptsize
\begin{tabular}{ll}
\toprule
\mbox{Latent dimension:} & $N_h=20$\\
\mbox{Parameters dimension:} & $N_{\text{par}}=14$\\
\mbox{Coupling layers:} & $N_f = 30$\\
\mbox{Masking strategy:} & $N_h/2$ frozen components (alternating)\\
\mbox{Dropout rate:} & $0.2$\\
\mbox{Weight initializer:} & \mbox{Xavier}\\
\mbox{$L^2$ regularization:} & $\lambda = 10^{-3}$\\
\mbox{Optimizer:} & Adam\\
\mbox{Batch size:} & $B_\text{NF} = 32$\\
\mbox{Initial learning rate:} & $10^{-3}$\\
\mbox{Max epochs:} & $200$\\
\mbox{Learning schedule:} & $0.8$ cosine decay\\
\mbox{Weight decay:} & \mbox{$0.05$}\\
\mbox{Early stopping patience:} & $20$ epochs\\
\mbox{Train--validation split:} & $80{:}20$\\
\bottomrule
\end{tabular}
}
\end{table}

\end{document}